# MAT: A simple yet strong baseline for identifying self-admitted technical debt

Zhaoqiang Guo, Shiran Liu, Jinping Liu, Yanhui Li, Lin Chen, Hongmin Lu, Yuming Zhou, Baowen Xu

State Key Laboratory for Novel Software Technology, Nanjing University
Department of Computer Science and Technology, Nanjing University

## Abstract

In the process of software evolution, developers often sacrifice the long-term code quality to satisfy the short-term goals due to specific reasons, which is called technical debt. In particular, self-admitted technical debt (SATD) refers to those that were intentionally introduced and remarked by code comments. Those technical debts reduce the quality of software and increase the cost of subsequent software maintenance. Therefore, it is necessary to find out and resolve these debts in time. Recently, many approaches have been proposed to identify SATD. However, those approaches either have a low accuracy or are complex to implementation in practice. In this paper, we propose a simple unsupervised baseline approach that fuzzily matches task annotation tags (MAT) to identify SATD. MAT does not need any training data to build a prediction model. Instead, MAT only examines whether any of four task tags (i.e. TODO, FIXME, XXX, and HACK) appears in the comments of a target project to identify SATD. In this sense, MAT is a natural baseline approach, which has a good understandability, in SATD identification. In order to evaluate the usefulness of MAT, we use 10 open-source projects to conduct the experiment. The experimental results reveal that MAT has a surprisingly excellent performance for SATD identification compared with the state-of-the-art approaches. As such, we suggest that, in the future SATD identification studies, MAT should be considered as an easy-to-implement baseline to which any new approach should be compared against to demonstrate its usefulness.

## 1 Introduction

Technical debt (TD) is a useful metaphor proposed by Cunningham [11] in 1993 to describe the situation that developers neglect the maintenance of long-term code quality in order to achieve short-term goals. For a software project, writing high quality and well-structured code is the initial goal. In the process of software evolution, however, uncertainties (e.g., restricted human, release time pressure, and cost reduction) [15, 16] often arise to interrupt the scheduled development plans. To cope with these changes and complete the previously designated tasks timely, developers are forced to adjust delivery projects with sub-optimal selection (e.g., hard-coded parameters and functional reduction) [17]. This solution temporarily solves the problem, but in the long term, it results in a chaotic structure and increases the effort of refactoring code [14, 30].

In particular, there is a kind of technical debt that is intentionally introduced (e.g., temporarily modifying a parameter) by developers and marked in code comments. Potdar et al. called the technical debts marked in code comments self-admitted technical debt (SATD) [14]. Their study indicated that SATD was common and might bring a negative impact to software maintenance. Also, Wehaibi et al. [15] conducted an empirical study to examine the relationship between SATD and software quality. Their findings showed that SATD not only might lead to software defects but also might make the software system more difficult to change in the future. Therefore, there is an urgent need to identify SATD and fix them timely.

Potdar et al. [14] first studied SATD according to code comments and summarized SATD comment patterns to identify SATD. In total, they manually read through 101762 comments and identified more than 60 recurring SATD patterns in four projects (i.e., Eclipse, Chromiun OS, ArgoUML, and Apache). Subsequently, Wehaibi et al. [15] and Bavota et al. [16] used the above 62 patterns to identify SATD for other projects. In particular, Bavota et al. [16] found that these patterns could identify SATD with a high precision. However, the pattern-based identification approach may have a low recall. The main reason for this is that the number of the identified patterns is limited so that cannot represent all possible situations.

In order to reduce manual effort of the pattern-based approach, many supervised approaches have been proposed in recent years [10]. First, Maldonado et al. [17] proposed an approach based on natural language processing (NLP) to automatically identify design and requirement SATD. Note that their approach was based on the Stanford Classifier [60] and showed a good performance. Subsequently, Huang et al. [10] proposed to use a text-mining based (TM) supervised approach to automatically identifying SATD. For each source project, they firstly trained a sub-classifier based on the feature data and the label information (i.e. SATD or not SATD) of the comments. After that, they used the sub-classifiers from multiple source projects to build a composite classifier. For a new comment from the target project, the composite classifier predicted its label by a voting strategy. Their results showed that the proposed approach considerably outperformed both the pattern-based approach and NLP approaches in terms of F1. Note that, the modeling process of the TM approach is relatively complex and it is difficult to explain the classification results. Most recently, Ren et al. [49] proposed a convolutional neural network (CNN) based model that not only had an excellent SATD classification performance, but also could extract explicable patterns (i.e., keywords or phrases). Nevertheless, as supervised approaches, the performance of NLP, TM, and CNN heavily depends on the quality and quantity of training data. Poor training data can lead to a low performance for all the three supervised approaches. In particular, if the training data is not available, these approaches cannot be applied in practice.

In this paper, we propose a simple unsupervised approach that fuzzily matches task annotation tags (MAT) to identify SATD. According to Storey et al.'s study [22], task annotation is important for individual, team, and community to use for software maintenance. In practice, task annotation tags have been used in many projects. Intuitively, if a source code comment contains such a task annotation tag, it is clearly that there is a SATD. Inspired by this intuition, we first study the usage of task tags and then select four representative task tags (i.e. TODO, FIXME, XXX, and HACK) from seven popular IDEs. Then, we apply a fuzzy match strategy to match task tags in code comments to identify SATD. Unlike Maldonado et al.'s [17], Huang et al.'s [10], and Ren et al's [49] supervised approaches, MAT does not need any training data to build a prediction model. Indeed, MAT only examines whether any of four task tags appears in the comments of a target project to identify SATD. Therefore, MAT is very simple to understand and is very easy to implement in practice. In order to evaluate the usefulness of MAT, we use 10 open-source projects collected by Maldonado et al. [17] conduct the experiment. The experimental results show that MAT shows a surprisingly strong performance in the identification of SATD compared with the state-of-the-art approaches. Given its simplicity and effectiveness, we strongly suggest MAT should be considered as an easy-to-implement baseline approach to which any new approach should be compared against to demonstrate its usefulness in future SATD identification studies.

In this paper, we make the following contributions:

- We propose an unsupervised approach MAT to identify SATD without involving any training data. In particular, MAT is very easy to understand and apply in the real-world projects.
- We conduct an extensive study to investigate the usefulness of MAT in practice. Based on ten open-source projects, the results show that MAT has a surprisingly strong performance compared with the state-of-the-art approaches.
- We provide the source code for implementing MAT, which can be easily used in future SATD Identification studies. This will facilitate researchers to use MAT as a baseline SATD identification approach, thus helping filter out actually useless approaches and develop really effective approaches.

The rest of this paper is organized as follows. Section 2 introduces the background of the state-of-the-art approaches for self-admitted technical debt identification. Section 3 presents the motivation of our approach. Based on this, we conduct some empirical studies in Section 4 and propose MAT, our task tag matching approach, in Section 5. Later, Section 6 describes the experimental setup and Section 7 reports the experimental results. After that, we discuss our approach and show the implications in Section 8. In Section 9, we introduce related research works. Section 10 analyzes the threats to the validity of our study. Section 11 concludes the paper and outlines the direction for our future work.

## 2 Self-admitted Technical Debt Identification

This section introduces four state-of-the-art approaches (including one unsupervised and three supervised approaches) for self-admitted technical debt identification.

### 2.1 Pattern matching based approach (Pattern)

Potdar et al. first studied SATD according to code comments. They summarized SATD comment patterns manually and proposed a pattern matching approach to identify SATD [14]. They manually read through 101762 source code comments and summarized more than 60 SATD comment patterns in four projects (i.e., Eclipse, Chromiun OS, ArgoUML, and Apache). The identified patterns listed in appendix A (actually, there are 63 patterns according to Huang et al.'s report [10]). Every pattern is a keyword (e.g., "hack", "ugly", and "stupid") or a phrase (e.g., "give up", "at a loss", and "get rid of this") that frequently appears in comments. The rationale of their approach is that a comment is considered to indicate SATD if one of the patterns appears in the comment. Therefore, pattern matching approach is an unsupervised approach that can be used directly to identify SATD. Besides, another advantage of the approach is that the patterns they summarized are intuitively understandable. According to Huang et al.'s [10] experimental result on eight open-source projects, Potdar et al.'s pattern matching approach achieved an excellent performance in precision (0.770 on average). However, the process of summarizing patterns will take up many human resources when handling a large number of comments. Furthermore, Huang et al. reported that Potdar et al.'s pattern matching approach had a very low recall. In other words, the Pattern approach can only identify a very small portion of SATD. The main reasons are two-folds. First, the number of patterns identified by Potdar et al. is limited, i.e. they cannot cover all possible situations. Second, different projects have different comment styles, which may have an influence on the identification of SATD.

## 2.2 Natural language processing based approach (NLP)

In order to overcome the limitations of Potdar et al.'s Pattern approach, Maldonado et al. [17] proposed an approach based on natural language processing (NLP) to automatically identify design and requirement SATD. In their experiment, they collected datasets that contained the comments from ten open source Java projects. For the datasets, they filtered out the comments that were less likely to be classified as self-admitted technical debt by applying heuristics. After that, they manually classified the label (i.e., SATD or non-SATD) of each comment. Based on the datasets, they used a Java implementation of a maximum entropy classifier, Stanford Classifier [60], to identify SATD. In particular, for the prediction of comments in each target project, they used the data of comments and labels from other projects as the training data. Their results showed that the NLP based approach achieved a good performance even with a relatively small training dataset when identifying SATD.

## 2.3 Text mining based approach (TM)

In addition to NLP based approach, Huang et al. [10] proposed a text-mining based (TM) supervised approach to automatically identify SATD. Assume that there are n source projects and one target project, the text mining approach consists of the following two phases: model building phase and model prediction phase. At the model building phase, they built a sub-classifier based on the comments in each source project. By default, each sub-classifier was trained using Naïve Bayes Multinomial (NBM) [21] model. At the model prediction phase, they used a vote strategy to composite n sub-classifiers to jointly predict the label (SATD or non-SATD) of an unknown comment in the target project. To this end, source code comments were first preprocessed using a general NLP process (i.e., tokenization, stop-word removal, and stemming) to get the corresponding set of features (a feature is a processed word). Since there were many features (i.e., tokens in comments) in a project (e.g., there were 3660 features in ArgoUML), they used Information Gain (IG) feature selection technology [43] to select a useful subset of features to avoid curse-of-dimensionality problem. The default selection ratio was 10% (e.g., 366 features were selected in ArgoUML). Then, they converted comments in a project into Vector Space Model (VSM) [33] represented by a matrix. The rows and columns indicated the comments and features (i.e., distinct words occurring all comments) respectively. They used TF-IDF (i.e., term frequency and inverse document frequency) as the value of element in the matrix. In particular, Huang et al. used eight open-source projects to evaluate the effectiveness of their proposed text mining approach. Their experimental results showed that the proposed TM approach was superior to Potdar et al.'s Pattern approach. However, Huang et al.'s TM approach is a complex supervised approach, whose performance heavily depend on the quality and quantity of training data.

**2.4 Convolutional neural network based approach (CNN)**

Although the TM outperforms than Pattern and NLP approaches in identifying SATD comments, its prediction results are hard to explain since the features (i.e., tokens) TM used are less intuitive than human-summarized patterns. In addition, there is an improvement space for the accuracy of SATD determination. Based on this, Ren et al. first summarize five key issues in SATD classification, including variant term frequency, project uniqueness, variable length, semantic variation, and class-imbalanced comment data [49]. After that, they proposed a convolutional neural network (CNN) based approach to improve the accuracy of SATD comments determination. In their approach, they learned a domain-specific word-embedding [51] based on skip-gram model [50] of word2vec from comments. In the process, each word was represented by a vector containing semantic information. Then, they concatenated all vectors of the words in a comment to a matrix as the input data of the CNN model. They first used the data from source projects to train the classification model and then used the model to prediction the labels (i.e., SATD or non-SATD) of comments in a target project. According to their experimental results, the CNN approach outperforms the Pattern, NLP, and TM approaches in identifying SATD. In addition to a good classification performance, their approach can exploit the computational structure of CNNs to automatically identify key phrases and patterns in code comments that are most relevant to SATD. This mechanism can reduce the effort of manually summarizing patterns and extract a large number of useful patterns. On the whole, the CNN approach is the latest state-of-the-art approach that works well in performance, generalizability, and explainability. The main limitation is that the model is complex, as it contains many hyper parameters (e.g., the dimension size of word embedding and the number of filters [52]) needed to be carefully tuned.

# 3 Motivation

**3.1 The necessity of a simple-yet-effective baseline approach in SATD identification**

Task tags are some indicative words. In recent years, many researchers have highlighted the necessity of a simple yet effective baseline approach in their respective research fields [61]. According to the latest studies [63], the benefits of using a simple yet effective baseline approach are mainly two-folds. On the one hand, this would ensure researchers to adequately compare and evaluate the performance of a newly proposed approach (SATD identification approach in our context). To be useful, a newly proposed approach should have a significantly better performance than the simple baseline approach and the corresponding effect size should be non-trivial. Otherwise, the motivation for introducing such a new approach could not be well justified. This is especially true if the proposed new approach is highly complex compared with the baseline approach. On the other hand, the "ongoing use of a baseline approach in the literature would give a single point of comparison" [63]. In the literature, a newly proposed approach is often compared against the state-of-the-art approaches. The underlying assumption is that a new approach is useful if it shows a superior performance. However, in our community, it is not a common practice to share their codes. As a result, researchers often have to re-implement the state-of-art approaches, where a tiny difference in the implementation may lead to a degraded performance. In this case, it will be misleading to report that the proposed new approach advances the state-of-the-art. Furthermore, due to the lack of a common baseline approach, it is not clear how far have we really progressed in the journey. These problems can be avoided if there is an ongoing use of a baseline approach whose code is publically available. In other words, a baseline approach defines a meaningful point of reference and hence allows a meaningful evaluation of any new approach against previous approaches.

The importance of a baseline approach has been well recognized in our research community. In the field of software engineering [61-64], many baseline approaches have been proposed. For example, Krishna et al. proposed a baseline approach for transfer learning [61]. Chen et al. used "sampling" as a baseline optimizer for search-based software engineering [62]. Zhou et al. suggested that ManualDown and ManualUp should be used as the baseline approaches in cross-project defect prediction [64]. In addition to software engineering, a large number of baseline approaches [65-73] also were proposed in many other fields. For instance, Xiao et al. proposed simple baselines for human pose estimation and tracking [69]. Ethayarajh used a strong but simple baseline to build unsupervised random walk sentence embedding [70].

**Table 1.** The comparison among state-of-the-art approaches

| Approach | Advantage | Disadvantage |
|---|---|---|
| Pattern | Simple and intuitive | Human-summarized and low accuracy |
| NLP | High accuracy | Complex model |
| TM | High accuracy | Less intuitive and complex model |
| CNN | High accuracy and intuitive | Complex model |

In SATD identification, there is no simple-yet-effective baseline approach available. Although a variety of approaches have been proposed (i.e. the pattern, NLP, TM, and CNN approaches), they do not satisfy the characteristics that a baseline approach should have. According to Whigham et al. [63] and Krishna et al. [61], in order to be both useful and widely used, a baseline approach at least should have the following important characteristics. First, it should be simple to describe, implement, and interpret. Second, it should be deterministic in its outcomes. Third, it should be publically available via a reference implementation and associated environment for execution. Fourth, it should offer a comparable performance to standard approaches. A baseline approach holding the above characteristics will facilitate researchers filter out actually useless approaches and to determine really useful approaches in SATD identification. As shown in Table 1, the Pattern approach has a low accuracy, while the NLP, TM, and CNN approaches are complex (many parameters need to be tuned). Currently, there is a need to develop a simple-yet-effective baseline approach for SATD identification.

## 3.2 The possibility to build a simple-yet-effective baseline approach in SATD identification

As Ren et al. stated, the essence of identifying SATD comments is a task of binary text classification [49]. The solutions to this kind of task usually depend on the extraction of semantic information and the understanding of context. As can be seen, the current state-of-the-art approaches described in Section 2 utilize the semantic information of comment texts in different ways. The Pattern approach summarized patterns manually to represent the characteristics of SATD. The NLP approach extracted the n-gram information to represent the semantic of sentences. The TM approach extracted top features as semantic context of SATD to train the classification model based on text-mining techniques. The CNN approach utilized the semantic word-embedding vectors, which contained rich semantic information, of words in comments to train a powerful prediction model. Table 1 summarizes the comparison among these state-of-the-art approaches. The unsupervised Pattern approach is the most intuitive and easiest to understand. However, its recall rate is too low. The other three supervised approaches can achieve a high accuracy but their modeling process is complex. In particular, there is no intuitive explanation for the given result predicted by the TM approach.

Is it possible to build a simple-yet-effective baseline approach for SATD identification? In order to answer this problem, we manually inspect the data sets used by previous studies [10, 17, 49]. Surprisingly, we find that many comments contain task annotation tags such as "TODO" and "FIXME" added by the programmer himself/herself. According to [5], such task tags are used as "reminders of actions, work to do or any other action required by the programmer". This means that they are strong indicators of SATD. Given this context, we have the following conjunction: if we use task tags to identify SATD, a high accuracy will be obtained. In other words, intuitively, it is highly possible to build a simple-yet-effective baseline approach to identify SATD by matching task tags in comments. When looking at the experimental results by the NLP, TM, and CNN approaches, we find that all of them consider these task tags as strong SATD indicators. In fact, such knowledge is prior, i.e. there is no need to study such knowledge by a learner. This intuition motivates us to use prior task tag knowledge to develop a baseline approach and to explore its advantages and disadvantages compared with the state-of-the-art approaches.

# 4 Empirical Study

Based on the above motivation, we next conduct a study to investigate the feasibility of building a simple-yet-effective baseline. First, we analyze the usefulness of task tags. Then, we summarize the representative task tags. Finally, we conduct a case study to explore the relationship between task tags and SATD comments.

## 4.1 The usefulness of task tags

Task tags are indicative words used as reminders for a work or action that needs to be done by a developer [2]. Usually, task tags are embedded in source code comments, i.e. a string followed by a short description. Well-known task tags are "TODO", "FIXME", and "XXX" [22]. Many popular IDEs (e.g., Eclipse[1] and NetBeans[2]) have supported developers to use task tags for team collaboration and task communication in development [58]. Fig. 1 presents the "preferences" window for configuring task tags in Eclipse. It can be seen that the above popular task tags have been predefined in the default settings.

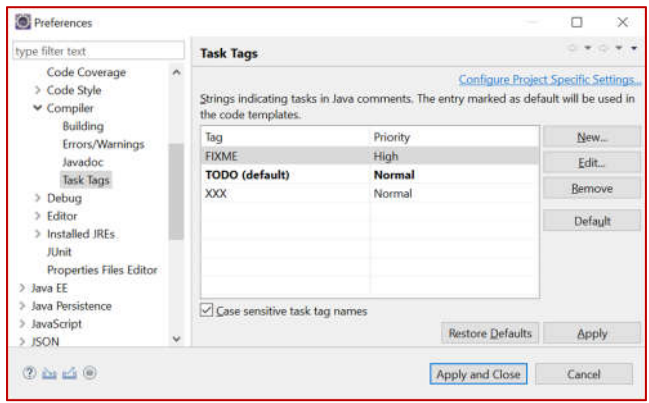

**Figure 1.** The preferences window for configuring task tags in Eclipse

Task tags are very useful in programming. For example, Fig. 2 presents a usage scenario of task tags (taken from [2]). When a developer are developing an application, he/she may define some methods (e.g. *printOddNumbers ()*) and decide to implement them later. To avoid forgetting to implement a method, the developer can add a task comment (//TODO implement the method *printOddNumbers ()*) to reminder the task. When checking the completeness of code, they can easily find out the code that needs to be implemented in time according to these (task) comments. In particular, Eclipse provides a predefined list of task tags that a developer can add to his/her code and view in a single location: the task view. A developer just needs to type a comment starting with the tag words "TODO" or "FIXME" in a new line. This line will automatically appear in the task window (Fig. 2 (b)) as soon as a developer saves the source code.

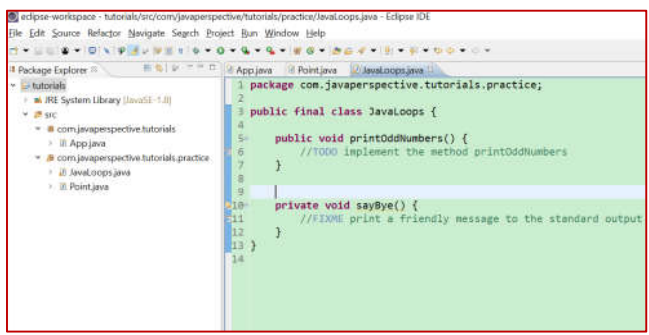

(a) The code comments that contain task tags

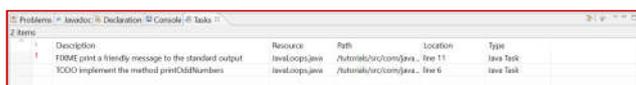

(b) The task list shown in the task window

---

[1] https://www.eclipse.org/
[2] https://netbeans.org/

**Figure 2.** A usage example of task tags in Eclipse

According to Storey et al. [22], task annotation tags play a key role in the work practices of software developers. In particular, task tags support articulation work such as the problem indicator and the edge case that would introduce underlying defects into a project. Similarly, task annotation tags in code comments would also occur near the problematic code. This inspires that task tags would be an indicator of SATD comment. In their paper, they have listed the popular task tags ("TODO", "FIXME", "XXX", and "HACK") and others.

**4.2 The representative task tags**

As a special kind of words, task tags usually use the words that can express the problem in a program. Developers can roughly understand what types of problems exist in one code by reviewing the task tag directly. After years of practice and development, many task tags have been gradually formed. Table 2 lists the default tags supported in popular IDEs. As can be seen, "TODO" is the most popular task tag used by all the seven tools. In addition, "FIXME", "XXX", and "HACK" are supported by at least two popular IDEs. In this sense, "TODO", "FIXME", "XXX", and "HACK" are the representative four task tags. For each of these four tags, Table 3 lists the corresponding meanings and use scenario [1]. In particular, Table 3 provides example uses of task tags in the comments of the projects collected by Maldonado et al. [17].

**Table 2.** The default task tags in popular IDEs (four representative tags are highlighted in bold)

| Popular IDEs | Default task tags |
|---|---|
| Eclipse [5] | **TODO,FIXME,XXX** |
| Visual Studio [7] | **TODO, HACK,** UNDONE, NOTE |
| IntelliJ IDEA [6] | **TODO, FIXME** |
| NetBeans [3] | @todo, **TODO, FIXME, XXX,** PENDING, <<<<<< |
| AndroidStudio [9] | **TODO, FIXME** |
| CodeClimate [8] | **TODO, FIXME, HACK, XXX,** BUG |
| Code::Blocks [4] | **TODO** |
| The representative tags | **TODO, FIXME, HACK, XXX** |

**Table 3.** The meanings and usage scenarios of four representative task tags

| Tags | Meanings and usage scenarios | Examples |
|---|---|---|
| TODO | Comments that mark something for later work, later revision or at least later reconsideration. TODO comments should be considered a very useful technique, although like all good things on Earth, there's certainly potential for abuse. | // TODO: Fully implement this! - [from ArgoUML]<br>// TODO implement clear() - [from Columba]<br>// TODO we should generate this. - [from EMF] |
| FIXME | A standard put in comments near a piece of code that is broken and needs work. Use FIXME to flag something that is bogus and broken. | // FIXME: not very efficient - [from JRuby]<br>// FIXME: There's some code duplication here... - [from JRuby] |
| XXX | A marker that attention is needed. Commonly used in program comments to indicate areas that are kluged up or need to be. Some hackers like `XXX' to the notional heavy-porn movie rating. Use it to flag something that is bogus but works. | // XXX - why not simply new File (dir, filename)? -[from Ant]<br>// XXX this should not hardcoded - [from jEdit] |
| HACK | Temporary code to force inflexible functionality, or simply a test change, or workaround a known problem. | // HACK: force the controller to load its tree - [from JMeter] |

**4.3 SATD comment case study**

To explore the characteristics of SATD comments, we conduct a case study based on the datasets provided by Maldomado et al. [17] (Huang et al. [10] and Ren et al. [49] also used the dataset in their experiment) to study the distribution of the above-mentioned four representative task tags in comments. In total, there are 10 projects in the datasets: Ant, ArgoUML, Columba, EMF, Hibernate, JEdit, JFreeChart, JMeter, JRuby, and SQuirrel. To reduce the manual effort, we randomly select 10% SATD comments and 10% non-SATD comments from each project to manually inspect. As a result, the numbers of the selected SATD comments range from 7 (EMF) to 96 (ArgoUML) on these projects, with a total of 277 SATD comments. The numbers of the selected non-SATD comments range from 211 (Hibernate) to 455 (ArgoUML) on these projects, with a total of 3420 non-SATD comments. Based on above sampled instances, we study the distribution of four representative task tags in comments.

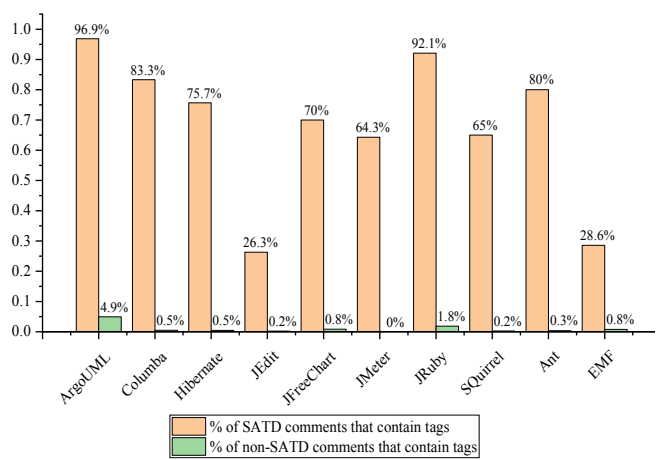

**Figure 3.** The percentage of instances that contain task annotation tags

Fig. 3 depicts the percentage of comments (including SATD comments and non-SATD comments) that contain representative task tags. According to the statistical results for SATD instances, the percentages of task tags embedded in comment instances are higher than 60% for 8 out of the 10 projects. In particular, for two projects (i.e., ArgoUML and JRuby), almost all the sampled instances include representative task tags. This indicates that it is common to use representative task tags to tag the code as suboptimal or remark underlying dangers. Note that, there are two special cases whose task tag inclusion ratio is lower than one-third (26% for JEdit and 29% for EMF). One possible reason is that the developers for these two projects may use their user-defined tags rather than the representative tags (shown in Table 2) we found. When looking at the statistical result for non-SATD instances, we can see that, the percentage of non-SATD instances that contain representative task tags is very low (less than 5%). This indicates that most non-SATD instances do not contain representative task tags.

**Table 4.** The detailed number of sampled instances that contain each task tag on each project

|  |  | ArgoUML | Columba | Hibernate | JEdit | JFreeChart | JMeter | JRuby | SQuirrel | Ant | EMF |
|---|---|---|---|---|---|---|---|---|---|---|---|
| SATD Comment | HACK | 0 | 0 | 0 | 2 | 0 | 2 | 4 | 1 | 1 | 0 |
|  | TODO | 92 | 8 | 28 | 1 | 4 | 16 | 21 | 12 | 3 | 2 |
|  | FIXME | 1 | 1 | 0 | 0 | 3 | 0 | 10 | 0 | 0 | 0 |
|  | XXX | 0 | 1 | 0 | 2 | 0 | 0 | 0 | 0 | 4 | 0 |
|  | #All Tags | 93 | 10 | 28 | 5 | 7 | 18 | 35 | 13 | 8 | 2 |
|  | #SATD | 96 | 12 | 37 | 19 | 10 | 28 | 38 | 20 | 10 | 7 |
| non-SATD Comment | HACK | 0 | 0 | 0 | 0 | 0 | 0 | 1 | 0 | 0 | 0 |
|  | TODO | 20 | 1 | 1 | 0 | 0 | 0 | 3 | 1 | 0 | 0 |
|  | FIXME | 0 | 1 | 0 | 0 | 2 | 0 | 2 | 0 | 0 | 0 |
|  | XXX | 2 | 0 | 0 | 1 | 0 | 0 | 0 | 0 | 1 | 2 |
|  | #All Tags | 22 | 2 | 1 | 1 | 2 | 0 | 6 | 1 | 1 | 2 |
|  | #non-SATD | 445 | 396 | 211 | 444 | 239 | 386 | 326 | 427 | 295 | 251 |

Table 4 reports the detailed number of sampled comments that contain each representative task tag on each project. As can been seen, "TODO" occurs in all the 10 projects for SATD comments and occupies most of the proportion (i.e., 85.39%) compared with the other tags. One possible reason is that many developers use auto-generated annotations by the IDE (e.g. Eclipse), which sets "TODO" as the default tag [5]. Therefore, "TODO" is the most popular tag for indicating SATD. The other tags such as "FIXME" and "XXX" have preferences on specified projects such JRuby and Ant but appear less frequently in the other projects. Considering the result for non-SATD comments, there are little task tags for non-SATD instances. Note that, some task tags (e.g., "HACK" and "XXX") do not occur in many projects.

In summary, we conclude that the four representative task tags (i.e. "TODO", "FIXME", "XXX", and "HACK") occur more frequently in SATD comments than in non-SATD comments. In other words, they should have a good ability to distinguish between SATD comments and non-SATD comments. Therefore, we may use four representative task tags to identify SATD.

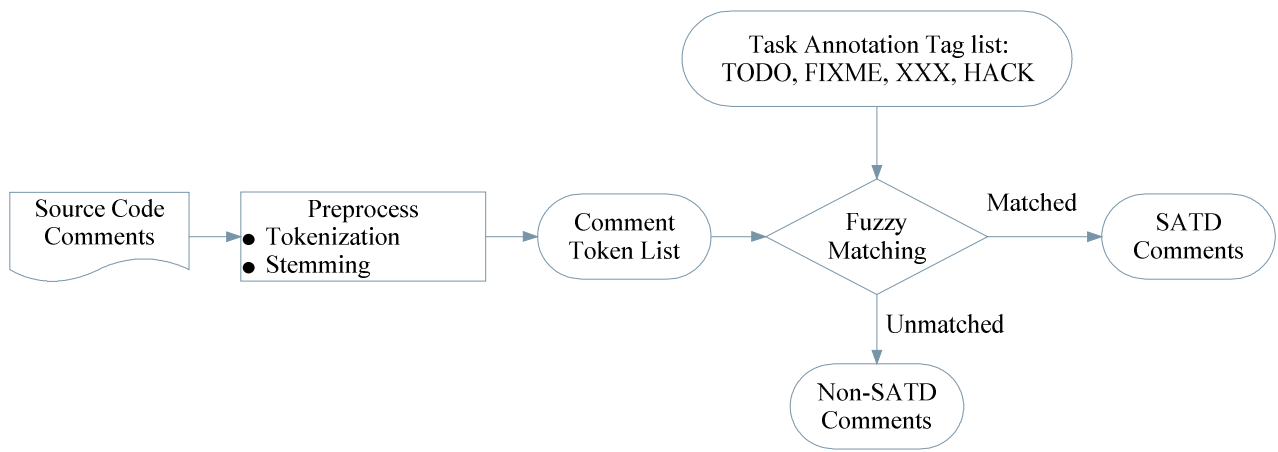

**Figure 4.** An overview of our MAT approach

## 5 MAT: Matching Annotation Tags

Inspired by the results in our empirical study, we develop a simple unsupervised approach that Matches the four representative task Annotation Tags (MAT) to identify SATD comments. In order to ensure the accuracy of our approach, we take the following two phases to complete the identification process: "preprocessing" and "fuzzy matching". Fig. 4 gives an overview of our approach.

1) **Preprocessing:** The first phase is to preprocess the text of comments. Different developers may have different habits of comment writing. For example, "TODO" can be written in the form of "todo", "Todo", or "TODO:". Therefore, we conduct a natural language preprocessing similar to that in Huang et al.'s study [10]. The "preprocessing" phase contains the following two main steps:

a) *Tokenization*: Divide the continuous text into single words and keep only English letters in a token.

b) *Stemming*: Transform every word into the original form (e.g., "hacks" to "hack") to improve the matching accuracy.

2) **Fuzzy matching:** The second phase is to match task tags in the preprocessed token list. A comment is considered indicating SATD if and only if there is at least one task tag ("TODO", "FIXME", "XXX", or "HACK") that occurs in the corresponding token list. At the same time, although the comments are preprocessed at the first phase, there are some unexpected cases that exist in the token set. For example, two words are linked together by deleting the space by a mistaken operation (e.g., "pleasefixme" and "hackhere"). These words cannot be directly matched with task tags. Therefore, we use a fuzzy matching strategy, which will match a word if a task tag is contained in the start or the end of a word, to find task tags. More specifically, we will consider tokens such as "fixme", "pleasefixme", or "fixmehere" as the matched tokens of "FIXME". If a comment consists of the above tokens, it will be classified as a SATD comment.

## 6 Experimental Setup

In this section, we introduce the experimental design to investigate the effectiveness of our proposed approach in identifying SATD, including the data sets, performance indicators, and the state-of-the-art approaches.

### 6.1 Data sets

In this study, we use the dataset shared by Huang et al. on their website[3]. On their website, there are two important data files: "comments" and "labels". The "comments" file lists the comments after filtering with heuristic rules from 10 open-source software projects, while the "labels" file lists the label (i.e. a SATD comment or a non-SATD comment) for each comment. These 10 open-source projects include ArgoUML 0.34, Columba 1.4, Hibernate 3.3.2, JEdit 4.2, JFreeChart 1.0.19, Jmeter 2.10, Jruby 1.4.0, Squirrel 3.0.3, Ant 1.7.0, and EMF 2.4.1. As stated by Huang et al., the dataset was originally provided by Maldonado et al [17, 18]. For each project, Maldonado et al. used an open-source Eclipse plug-in (i.e., JDeodrant [48]) to parse source code and extract code comments. In particular, they applied the following five heuristic rules to filter out those comments that were obviously impossible to be SATD comments: (1) removed license comments; (2) grouped consecutive single-line comments as one comment; (3) removed commented source code fragments that usually did not contain SATD; (4) removed automatically generated comments by the IDE (e.g., Eclipse); and (5) removed Javadoc comment unless they contained at least one task annotation tag (e.g., "TODO:" and "FIXME:"). After obtaining the comments filtered with these heuristic rules, Maldomado et al. manually determined their labels (i.e. whether a comment was a SATD comment).

---
[3] https://github.com/tkdsheep/TechnicalDebt

**Table 5.** The summary for the dataset used in our study

| Project-Release | Domain | #Comments | After filtering | #SATD | % of SATD | #feature | Cont. | KLOC |
|---|---|---|---|---|---|---|---|---|
| Ant-1.7.0 | Project Build Tool | 21587 | 3052 | 102 | 3.34% | 2589 | 74 | 115 |
| ArgoUML-0.34 | UML Modeling Tool | 67716 | 5426 | 969 | 17.86% | 3660 | 87 | 926 |
| Columba-1.4 | Email Client | 33895 | 4090 | 128 | 3.13% | 2676 | 10 | 155 |
| EMF-2.4.1 | Eclipse Modeling Framework | 25229 | 2585 | 74 | 2.86% | 2153 | 30 | 228 |
| Hibernate-3.3.2 | ORM Framework | 11630 | 2492 | 377 | 15.13% | 2418 | 314 | 703 |
| JEdit-4.2 | Text Editor | 16991 | 4644 | 195 | 4.20% | 3888 | 57 | 310 |
| JFreeChart-1.0.19 | Chart Library | 23474 | 2494 | 101 | 4.05% | 2012 | 19 | 317 |
| Jmeter-2.10 | Performance Tester | 20084 | 4148 | 282 | 6.80% | 2751 | 41 | 354 |
| Jruby-1.4.0 | Ruby Interpreter | 11149 | 3652 | 383 | 10.49% | 2891 | 374 | 841 |
| Squirrel-3.0.3 | SQL Client | 27474 | 4473 | 201 | 4.49% | 3329 | 40 | 708 |
| Average | | 25922.9 | 3705.6 | 281.2 | 7.23% | 2836.7 | 104.6 | 465.7 |
| Total | | 259229 | 37056 | 2812 | 7.59% | - | 1046 | 4657 |

Table 5 provides the summary for the dataset used in our study. The first column reports for each project the name and the corresponding version. The second column reports the application domain. The third and fourth columns report the number of original comments and number of filtered comments. The fifth and sixth columns report the number and the proportion of SATD comments in the filtered comments. The seventh column reports the number of features (i.e., distinct words) after text preprocessing. The eighth column reports for each project the number of contributors extracted from an online community and public directory OpenHub [47]. The last column reports the project size counted by SLOCCount [46]. On average, the number of comments for a project is about 3705 and 7.23% of them are SATD comments.

Note that, Huang et al. only used 8 out of 10 data sets to conduct their study (Ant 1.7.0 and EMF 2.4.1 were not used). Furthermore, we find that, for each of the 8 projects, the number of comments filtered with the heuristic rules on their website is slightly different from that in their published paper. After communicating with Huang, we were told that their dataset was updated after their paper was published. In other words, the dataset on their website is more reliable. Therefore, in our paper, we use the up-to-date dataset (from 10 projects) on their website to evaluate the effectiveness of MAT as well as the state-of-the-art approaches.

### 6.2 Performance indicators

In nature, the process of identifying SATD comments is a typical binary classification problem [42-44], which aims to determine whether a comment indicates SATD or not. For the classification result, there are total four situations (TP, FP, TN, and FN) that usually represented by the confusion matrix shown in Fig. 5. Each row represents the comments in a predicted class while each column represents the comments in an actual class. More specifically, TP (true positive) denotes the set of correctly classified SATD comments, FP (false positive) denotes the set of mistakenly classified non-SATD comments, TN (true negative) denotes the set of correctly classified non-SATD comments, and FN (false negative) denotes the set of mistakenly classified SATD comment.

|  |  | Actual label | |
|---|---|---|---|
|  |  | SATD | Non-SATD |
| Predicted label | SATD | TP | FP |
|  | Non-SATD | FN | TN |

**Figure 5.** The confusion matrix

Based on the aforementioned confusion matrix, we use the following three popular performance indicators to evaluate the classification performance of a SATD identification approach.

- **Precision** is the proportion of comments that are correctly classified as SATD comments among those classified as SATD comments. If the precision of an approach is high, the SATD comments identified by this approach are usually correct.

$$Precision = \frac{|TP|}{|TP| + |FP|}$$

- **Recall** is the proportion of comments that are correctly classified as SATD comments among those true SATD comments. If the recall of an approach is high, a large percentage of real SATD comments can be found in the result of classification.

$$Recall = \frac{|TP|}{|TP| + |FN|}$$

- **F1** is the harmonic mean of precision and recall. According to Han et al. [21], precision and recall are complementary indicators. That is, in many cases, the increase of the precision of an approach will lead to a decrease of the recall and vice versa. Therefore, it is not comprehensive to evaluate the performance using the two indicators separately. F1 reflects the comprehensive contribution of both.

$$F_1 = \frac{2 \times Precision \times Recall}{Precision + Recall}$$

## 6.3 State-of-the-art approaches

Fig. 6 shows the prediction settings in our study. In order to evaluate the effectiveness of MAT, we compare it with the following state-of-the-art approaches: Potdar et al.'s Pattern approach [14] (Pattern for short), Maldonado et al.'s natural language processing based approach (NLP for short [17]), Huang et al.'s text mining approach (TM for short [10]), and Ren et al.'s CNN approach [49]. As described in Section 2, the pattern approach is an unsupervised approach, while the NLP, TM, and CNN approaches are supervised approaches.

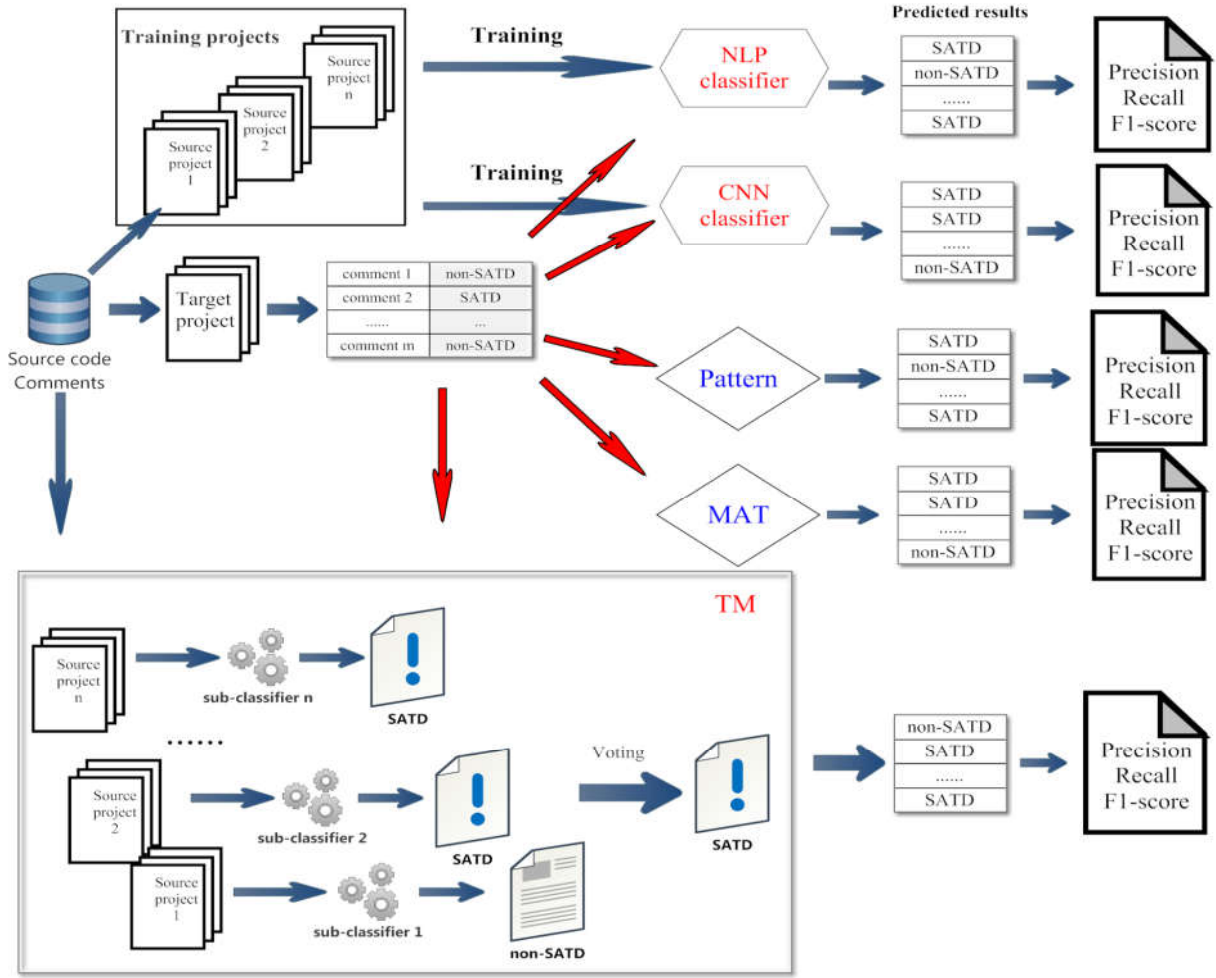

**Figure 6.** The prediction settings of five approaches

**Pattern.** We use the patterns (i.e. keywords or phrases appearing frequently in SATD comments, as shown in Appendix A) summarized by Potdar et al. to identify the label of each comment in the target project. If any of these patterns is found in a comment, it will be classified as a SATD comment (i.e. it indicates a SATD). Otherwise, it will be classified as a non SATD comment. Similar to the Pattern approach, our MAT approach is also an unsupervised approach. However, there are two important differences. On the one hand, MAT use only four task annotation tags to identify SATD, far less than the number of patterns in the Pattern approach. On the other hand, MAT takes a fuzzy matching strategy, while the Pattern approach takes a strict matching strategy.

**NLP.** Since Maldonado et al. [17] and Huang et al. [10] did not share the source code of the NLP approach, we read their papers carefully and implemented the NLP approach by ourselves to compare the performance with MAT. Note that, the core of NLP approach is based on the Stanford Classifier [60]. Our implementation also used the Stanford classifier to train the NLP model.

**TM.** We use the code shared by Huang et al. [10] to implement the TM approach. Specifically, for a given target project, we first train a single sub-classifier for each of the other 9 individual source projects. Then, we use the resulting 9 sub-classifiers to vote for the label of each comment in the target project.

**CNN.** The code of Ren et al.'s CNN approach is not shared online. In our study, we do not attempt to re-implement the CNN approach, as it is hard to reproduce the experimental results due to the involvement of many parameters that need to be carefully tuned. Since the datasets in our study are the same as those in Ren et al.'s study [49], we use the prediction performance values of CNN reported in their study to conduct the comparison. Therefore, the implementation bias can be avoided. This ensures that we make a fair comparison between the CNN approach and MAT in SATD identification.

## 6.4 Research questions

**RQ1: (Influence of fuzzy matching strategy)** *How does fuzzy matching strategy affect the classification effectiveness of MAT?*

*Motivation.* As aforementioned, comment words may be connected together (e.g., "pleasefixme", "hackhere") due to carelessness. These typos will negatively increase the proportion of mismatching when using a strict matching strategy. In order to tackle this problem, MAT takes a fuzzy matching strategy (as described in Section 5). To determine the usefulness of a fuzzy strategy, in this RQ, we compare the effectiveness of a strict matching strategy versus a fuzzy matching strategy.

*Approach.* We construct two MAT models that respectively apply strict and fuzzy strategies to match tags in comments for SATD identification. The only difference between the two models is the selection of matching strategy. For each target project, MAT will identify the label (i.e., SATD or non-SATD) of each comment in the project. After that, we have 10 classification results for the models that apply strict strategy and fuzzy strategy respectively. Based on the above results, we compute precision, recall, and F1 to evaluate the effectiveness of each matching strategy.

**RQ2: (Classification performance of MAT)** *How effective is MAT in identifying SATD compared with the state-of-the-art approaches?*

*Motivation.* As mentioned in previous sections, MAT is a very simple and easy-to-understand approach in identifying SATD. This is very helpful for practical use. Being an unsupervised approach, MAT can be applied to identify SATD directly in a target project without the need to train a model in advance. This is an obvious advantage over the (complex) supervised approaches. We want to know how well does MAT perform in classification performance compared with the state-of-the-art approaches. If MAT exhibits a very competitive or even superior classification performance, we then recommend that developers should apply the simple approach instead of using the complex approach in practice.

*Approach.* To answer this research question, we use the same data sets (from 10 projects) to evaluate the classification performance of MAT as well as the state-of-the-art approaches under the following two scenarios:

- *Scenario 1: Many-to-one (MTO) prediction among 10 projects.* Under this scenario, for each supervised approach (i.e., NLP, TM, and CNN), each data set is used as the test data set to evaluate a model built on the training data set combined from the remaining 9 data sets. For each unsupervised approach (i.e. the Pattern approach and the MAT approach), it is directly applied to the test data without a training process. For each approach, we have 10 experiments. In total, we have 50 classification results (5 approaches × 10 projects). After obtaining the results on the 10 projects, we can investigate the classification effectiveness (i.e. precision, recall, and F1) of MAT compared with the Pattern, NLP, TM, and CNN approaches.

- ***Scenario 2: One-to-One (OTO) prediction among 10 projects.*** Under this scenario, on a given testing data set, for each supervised approach (i.e., NLP, TM, and CNN), we have 9 experiments (i.e., we train 9 models using 9 source projects to predict the labels of comments in target project) and use their average performance as the resulting performance. For each unsupervised approach, we have 1 experiment (i.e., we predict the labels of comment in target project directly) on each testing data set. In total, we have 50 classification results (5 approaches × 10 projects). After obtaining the results on 10 projects, we can investigate the effectiveness of MAT compared with the Pattern, NLP, TM, and CNN approaches.

Scenario 1 aims to simulate the scenario that rich training data are available, while scenario 2 aims to simulate the scenario that only limited training data are available. Under each scenario, we employ the Wilcoxon-signed rank test to examine whether there is a statistically significant difference between MAT and each of the state-of-the-art approaches at the significance level of 0.05. Furthermore, we use the Cliff's delta ($\delta$), which is used for median comparison, to quantify the effect size of the difference. This enables us to know whether the magnitude of the difference between the performances of two approaches is important from the viewpoint of practical application. By convention, the magnitude of the difference is considered trivial ($|\delta| < 0.147$), small (0.147-0.33), moderate (0.33-0.474), or large ($> 0.474$).

**RQ3: (Difference in correct classification results)** *What is the difference of the correct classification results between MAT and other approaches in SATD identification?*

*Motivation.* In RQ2, we calculate an overall performance score (such as precision, recall, and F1) to represent how well an approach correctly predicts SATD by taking into account the level of incorrect predictions made. However, only the overall performance score cannot illustrate the classification result of single comment classified by different approaches. According to [74], the overall performance can hide a variety of differences in the defects that each classifier predicts. Similarly, for the problem of SATD classification, it is important to investigate the difference in the individual SATD and non-SATD identified by various approaches. Studying the difference will help us understand the classification characteristics of various approaches such as the overlapped instances and unique instances classified by different approaches. As a result, we can analyze the advantages and disadvantages of each approach. More specifically, in our study, we investigate the difference of the correct classification results among MAT, the NLP approach, and the TM approach. Note that, in RQ3, we do not take into account the Pattern approach and the CNN approach. The reasons are in the following: for the former approach, the recall is too low to conduct the analysis; for the latter approach, its code is not shared online (it is hard to implement it by ourselves to exactly reproduce Ren et al.'s results due to many parameters involved).

*Approach.* According to Maldonado et al.'s study [17] and Huang et al.'s study [10], both the NLP and TM approaches were evaluated under the MTO scenario. Consistent with their studies, we investigate RQ3 under the MTO scenario. First, for each project, we obtain the true SATD comments (i.e. true positive SATD comments) that three approaches (i.e. the MAT, NLP, and TM approaches) can identify. Second, we use a set diagram to describe the difference between three approaches in terms of the specific SATD comments that each approach identifies and does not identify. In order to make an in-depth understanding of the difference between three approaches, we apply a similar analysis to the true non-SATD comments (i.e. true negative non-SATD comments) that three approaches can identify.

# 7 Experimental Results

In this section, we report the experimental results in detail. To enable external replication, we make the data sets, source codes of MAT, and experimental results in our study publicly available[4].

## 7.1 RQ1: Influence of fuzzy matching strategy

*Result.* Table 6 reports the performance comparison between strict and fuzzy matching strategy in terms of precision, recall, and F1. The improvement rates marked by bold fonts indicate that the performance of fuzzy strategy is superior to that of strict strategy. As can be seen, on average, the fuzzy strategy leads to a slightly higher recall and an almost same precision compared with the strict strategy. In other words, the fuzzy strategy can identify more SATD comments and at the same time it almost does not introduce false positives. Note that, the recalls of more than half of the projects (6/10) have increased, which indicates that there are typos in the comments of these projects and the occurrence of typos is not accident. Therefore, it is necessary to use fuzzy matching strategy to identify these SATD comments. In terms of F1, fuzzy matching achieves values has an average of 0.722, which is higher than the F1 of strict matching (0.716). In comparison, fuzzy strategy slightly improves F1 for 6 out of the 10 projects and leads to an inferior F1 in only one project. On average, fuzzy strategy leads to an improvement of 0.8% in terms of F1.

Table 6. Performance comparisons between strict and fuzzy matching strategy

| Projects | Precision | | | Recall | | | F1 | | |
|---|---|---|---|---|---|---|---|---|---|
| | strict | fuzzy | Improv. | strict | fuzzy | Improv. | strict | fuzzy | Improv. |
| Ant | 0.865 | 0.870 | **0.6%** | 0.441 | 0.461 | **4.5%** | 0.584 | 0.603 | **3.3%** |
| ArgoUML | 0.838 | 0.823 | -1.8% | 0.934 | 0.934 | 0.0% | 0.883 | 0.874 | -1.0% |
| Columba | 0.912 | 0.906 | -0.7% | 0.813 | 0.828 | **1.8%** | 0.860 | 0.865 | **0.6%** |
| EMF | 1.000 | 1.000 | 0.0% | 0.338 | 0.351 | **3.8%** | 0.505 | 0.520 | **3.0%** |
| Hibernate | 0.944 | 0.945 | **0.1%** | 0.714 | 0.724 | **1.4%** | 0.813 | 0.820 | **0.9%** |
| JEdit | 0.844 | 0.851 | **0.8%** | 0.195 | 0.205 | **5.1%** | 0.317 | 0.331 | **4.4%** |
| JFreeChart | 0.723 | 0.723 | 0.0% | 0.723 | 0.723 | 0.0% | 0.723 | 0.723 | 0.0% |
| JMeter | 0.924 | 0.924 | 0.0% | 0.780 | 0.780 | 0.0% | 0.846 | 0.846 | 0.0% |
| JRuby | 0.911 | 0.911 | 0.0% | 0.877 | 0.883 | **0.7%** | 0.894 | 0.897 | **0.3%** |
| SQuirrel | 0.925 | 0.925 | 0.0% | 0.612 | 0.612 | 0.0% | 0.737 | 0.737 | 0.0% |
| Average | 0.889 | 0.888 | -0.1% | 0.643 | 0.650 | **1.1%** | 0.716 | 0.722 | **0.8%** |

*Conclusion.* In summary, by correctly dealing with the typos of task tags, fuzzy matching strategy can improve the recall and maintain almost the same precision compared with strict matching strategy. This can facilitate developers to find more SATD comments by MAT in practice.

## 7.2 RQ2: Classification performance of MAT

---

[4] https://github.com/Naplues/MAT

***Result under the MTO scenario.*** Table 7-9 report the performance comparison results for under the MTO prediction scenario. For each table, we list the performance scores of each approach and the improvement rates that MAT achieves compared with each of the state-of-the-art approaches. In particular, we use bold fonts to highlight the improvement rates that are larger than 0%. In addition, as can be seen, on average, MAT can achieve a precision of 0.888, a recall of 0.650, and a F1 value of 0.722. This result is excellent for SATD comments identification. However, there are some exceptional cases. For instance, MAT has a recall lower than 0.500 in three projects (i.e., 0.205 for JEdit, 0.461 for Ant, and 0.351 for EMF). The reason for this is that the four task tags used in MAT rarely exist in these projects. According to the results in Table 7-9, we have the following observations:

- MAT vs. Pattern. In terms of precision, MAT has a better (higher) value on 8 out of 10 projects. Considering the recall and F1 of MAT, it has an average improvement of 827.25% and 479.92% compared with Pattern respectively. The great improvement on recall indicates that MAT can identify more SATD comments than Pattern. This is mainly caused by the inherent imprecision of text matching via limited patterns (i.e., keywords or phrase). In other words, natural language is more diversified so that a fixed pattern cannot match another expression of that pattern. For example, there is a pattern called *"remove this code"* according to Potdar et al. However, a SATD comment *"todo remove this old implementation after it's demonstrated that it's not needed"* (from ArgoUML project) cannot be identified by the Pattern even though their semantics are very similar. Because of this reason, the Pattern approach may leave out many of SATD comments.
- MAT vs. NLP. In terms of precision, MAT has an average improvement of 30.12% compared with NLP. Note that the precision of MAT is higher than that of NLP in all target projects. When looking at recall, we find that MAT can identify more SATD comments in the majority (6 out of 10) projects. In terms of F1, on average, MAT has an improvement of 10.35% compared with NLP.
- MAT vs. TM. Compared with TM, on average, MAT has an improvement of 16.72% and 3.71% in terms of precision and F1. In particular, for precision, MAT is better than TM in all projects. Meanwhile, the average recall of MAT is slightly lower than that of TM (0.650 vs. 0.655).
- MAT vs. CNN. Compared with CNN, we find that MAT can achieve an average improvement of 12.61% in terms of precision. Note that, the precision of MAT is higher in all target projects. In particular, MAT can achieve a greatly improvement on many projects (e.g., 48.97% for Ant). As an unsupervised approach, this is a great advantage for precise identification of SATD comments. The comparison between MAT and CNN in terms of recall shows that MAT can identify less SATD comments than CNN. It is easy to understand that some of SATD comments do not marked by task tags so that they cannot be identified by MAT. However, CNN can identify these comments by semantic analysis. The situation is particularly evident for JEdit, on which MAT decreases the recall value of 58.08% compared with CNN. According to the F1 scores, MAT achieves a value of 0.722 that is slightly lower than that (0.764) of CNN.

**Table 7.** The performance comparison under the MTO scenario in terms of precision

| Projects | Pattern | Impr.% | NLP | Impr.% | TM | Impr.% | CNN | Impr.% | MAT |
|---|---|---|---|---|---|---|---|---|---|
| Ant | 0.556 | 56.47% | 0.476 | 82.77% | 0.565 | 53.98% | 0.584 | 48.97% | 0.870 |
| ArgoUML | 0.743 | 10.77% | 0.798 | 3.13% | 0.816 | 0.86% | 0.816 | 0.86% | 0.823 |
| Columba | 0.900 | 0.67% | 0.754 | 20.16% | 0.817 | 10.89% | 0.830 | 9.16% | 0.906 |
| EMF | 0.500 | 100.00% | 0.433 | 130.95% | 0.636 | 57.23% | 0.793 | 26.10% | 1.000 |
| Hiberbate | 0.900 | 5.00% | 0.822 | 14.96% | 0.876 | 7.88% | 0.930 | 1.61% | 0.945 |
| JEdit | 0.857 | -0.70% | 0.667 | 27.59% | 0.772 | 10.23% | 0.773 | 10.09% | 0.851 |
| JFreeChart | 0.833 | -13.21% | 0.663 | 9.05% | 0.627 | 15.31% | 0.686 | 5.39% | 0.723 |
| JMeter | 0.778 | 18.77% | 0.757 | 22.06% | 0.858 | 7.69% | 0.873 | 5.84% | 0.924 |
| JRuby | 0.750 | 21.47% | 0.796 | 14.45% | 0.855 | 6.55% | 0.805 | 13.17% | 0.911 |
| SQuirrel | 0.471 | 96.39% | 0.657 | 40.79% | 0.784 | 17.98% | 0.794 | 16.50% | 0.925 |
| Average | 0.729 | 21.84% | 0.682 | 30.12% | 0.761 | 16.72% | 0.788 | 12.61% | 0.888 |

**Table 8.** The performance comparison under the MTO scenario in terms of recall

| Projects | Pattern | Impr.% | NLP | Impr.% | TM | Impr.% | CNN | Impr.% | MAT |
|---|---|---|---|---|---|---|---|---|---|
| Ant | 0.098 | 370.41% | 0.480 | -3.96% | 0.471 | -2.12% | 0.758 | -39.18% | 0.461 |
| ArgoUML | 0.027 | 3355.56% | 0.878 | 6.26% | 0.856 | 9.00% | 0.950 | -1.79% | 0.933 |
| Columba | 0.070 | 1082.86% | 0.742 | 11.59% | 0.805 | 2.86% | 0.875 | -5.37% | 0.828 |
| EMF | 0.054 | 550.00% | 0.392 | -10.46% | 0.473 | -25.79% | 0.594 | -40.91% | 0.351 |
| Hiberbate | 0.072 | 905.56% | 0.674 | 7.42% | 0.729 | -0.69% | 0.743 | -2.56% | 0.724 |
| JEdit | 0.185 | 10.81% | 0.369 | -44.44% | 0.364 | -43.68% | 0.489 | -58.08% | 0.205 |
| JFreeChart | 0.050 | 1346.00% | 0.644 | 12.27% | 0.733 | -1.36% | 0.802 | -9.85% | 0.723 |
| JMeter | 0.050 | 1460.00% | 0.706 | 10.48% | 0.770 | 1.30% | 0.787 | -0.89% | 0.780 |
| JRuby | 0.055 | 1505.45% | 0.794 | 11.21% | 0.752 | 17.42% | 0.930 | -5.05% | 0.883 |
| SQuirrel | 0.040 | 1430.00% | 0.687 | -10.92% | 0.597 | 2.51% | 0.692 | -11.56% | 0.612 |
| Average | 0.070 | 827.25% | 0.637 | 2.10% | 0.655 | -0.76% | 0.762 | -14.70% | 0.650 |

Table 10 reports the statistical test results under the MTO prediction scenario. In the view of statistical test, we make the following observations. First, MAT is significantly better than Pattern (the effect size is large), regardless of whether precision, recall, or F1 is considered. Second, when compared with NLP and TM, MAT exhibits a significantly better precision (the effect size is large), although their difference in F1/recall is not significant. Third, the precision of MAT is significantly better than CNN (the effect size is large) while the recall of MAT is significantly lower than CNN (the effect size is small, yet). Considering the F1 score, there is no significant difference between them. The above results clearly reveal that MAT performs better than Pattern, NLP, and TM in identifying SATD comments. Meanwhile, being an unsupervised approach, MAT is competitive to the supervised approach CNN. As a result, for practitioners, it would be better to apply MAT to locate SATD in a target project.

Combining the above results, we find that, under the MTO prediction scenario: MAT has an outstanding classification performance compared with the unsupervised approach Pattern. In addition, MAT is very competitive to the supervised approaches (NLP, TM, and CNN) in SATD identification.

Table 9. The performance comparison under the MTO scenario in terms of $F_1$

| Projects | Pattern | Impr.% | NLP | Impr.% | TM | Impr.% | CNN | Impr.% | MAT |
|---|---|---|---|---|---|---|---|---|---|
| Ant | 0.167 | **261.08%** | 0.478 | **26.15%** | 0.513 | 17.54% | 0.660 | -8.64% | 0.603 |
| ArgoUML | 0.052 | **1580.77%** | 0.836 | **4.55%** | 0.835 | **4.67%** | 0.878 | -0.46% | 0.874 |
| Columba | 0.130 | **565.38%** | 0.748 | **15.64%** | 0.811 | **6.66%** | 0.852 | **1.53%** | 0.865 |
| EMF | 0.098 | **430.61%** | 0.411 | **26.52%** | 0.543 | -4.24% | 0.679 | -23.42% | 0.520 |
| Hiberbate | 0.133 | **516.54%** | 0.741 | **10.66%** | 0.796 | **3.02%** | 0.826 | -0.73% | 0.820 |
| JEdit | 0.304 | **8.88%** | 0.475 | -30.32% | 0.495 | -33.13% | 0.599 | -44.74% | 0.331 |
| JFreeChart | 0.093 | **677.42%** | 0.653 | **10.72%** | 0.676 | **6.95%** | 0.739 | -2.17% | 0.723 |
| JMeter | 0.093 | **809.68%** | 0.730 | **15.89%** | 0.811 | **4.32%** | 0.828 | **2.17%** | 0.846 |
| JRuby | 0.102 | **779.41%** | 0.795 | **12.83%** | 0.800 | **12.13%** | 0.836 | **7.30%** | 0.897 |
| SQuirrel | 0.073 | **909.59%** | 0.672 | **9.67%** | 0.678 | **8.70%** | 0.739 | -0.27% | 0.737 |
| Average | 0.125 | **479.92%** | 0.654 | **10.35%** | 0.696 | **3.71%** | 0.764 | -5.50% | 0.722 |

Table 10. The results from the Wilcoxon-singed rank test and the Cliff's delta when comparing MAT with state-of-the-art approaches under the MTO scenario (bold fonts denote that the p-value is less than 0.05)

| | Indicators | Precision | Recall | F1 |
|---|---|---|---|---|
| p-value | Pattern | **0.037** | **0.002** | **0.002** |
| | NLP | **0.001** | 0.413 | 0.054 |
| | TM | **0.002** | 0.799 | 0.106 |
| | CNN | **0.002** | **0.006** | 0.375 |
| Cliff's delta (Effect size) | Pattern | 0.680 (large) | 1.000 (large) | 1.000 (large) |
| | NLP | 0.917 (large) | 0.107 (negligible) | 0.306 (small) |
| | TM | 0.720 (large) | 0.000 (negligible) | 0.280 (small) |
| | CNN | 0.620 (large) | -0.260 (small) | 0.040 (negligible) |

***Result under the OTO scenario.*** Table 11-13 report the performance comparison results under the OTO prediction scenario. Since MAT and Pattern are unsupervised approaches, they are not influenced by the training data. As a result, they have the same performance under the MTO and OTO scenarios. According to the results in Table 11-13, we have the following observations:

- MAT vs. NLP. In terms of precision, MAT has a better (higher) value on 8 out of 10 projects, which has an average improvement of 18.81%. Considering the recall, on average, MAT performs better than NLP in all 10 projects, which has an improvement of 37.95%. Similarly, we can see that, MAT achieves better F1 scores than NLP in all the 10 projects, with an average improvement of 29.74%. Note that, the average precision of NLP increases under the OTO prediction scenario (0.682 under MTO vs. 0.747 under OTO). Besides, the average F1 score under the OTO scenario (0.556) is slightly lower than that under the MTO scenario (0.654). The above results show that NLP can achieve a good performance even with a relatively small training dataset (i.e., one training project) when identifying SATD comments [17].

- MAT vs. TM. Compare with TM, on average, MAT has an improvement of 111.23%, 3.08%, and 47.78% in terms of precision, recall, and F1 respectively. In particular, for precision, MAT is better than TM in all projects. We can see that, under the OTO scenario, the average precision of TM decreases largely (0.761 under MTO vs. 0.420 under OTO). This indicates that, the precision of TM is sensitive to training set size. Many false positive instances will be introduced by TM when training set is small. As for the average recall, there is no large difference under the two scenarios (0.655 under MTO vs. 0.631 under OTO), which means that the recall of TM is not sensitive to training set size.

- MAT vs. CNN. Compared with CNN, MAT can achieve an average improvement of 59.50% in terms of precision. Different from the MTO scenario, MAT achieves better an average recall than CNN under the OTO scenario, with an average improvement of 32.98%. Considering the F1 score, MAT performs better than CNN in most (8 out of 10) projects, with an average improvement of 28.13%. This means that MAT is better than the latest supervised approach when the training set is small. Note that, CNN achieves a low average recall (0.489) under the OTO scenario compared with that (0.762) under the MTO scenario. The large degradation reveals that many SATD comments cannot be identified by CNN when the training data are limited.

Table 11. The performance comparison under the OTO scenario in terms of precision

| Projects | Pattern | Impr.% | NLP | Impr.% | TM | Impr.% | CNN | Impr.% | MAT |
|---|---|---|---|---|---|---|---|---|---|
| Ant | 0.556 | 56.47% | 0.560 | 55.25% | 0.183 | 375.41% | 0.440 | 97.73% | 0.870 |
| ArgoUML | 0.743 | 10.77% | 0.824 | -0.08% | 0.680 | 21.03% | 0.600 | 37.17% | 0.823 |
| Columba | 0.900 | 0.67% | 0.809 | 12.02% | 0.413 | 119.37% | 0.525 | 72.57% | 0.906 |
| EMF | 0.500 | 100.00% | 0.566 | 76.83% | 0.199 | 402.51% | 0.512 | 95.31% | 1.000 |
| Hiberbate | 0.900 | 5.00% | 0.895 | 5.53% | 0.622 | 51.93% | 0.544 | 73.71% | 0.945 |
| JEdit | 0.857 | -0.70% | 0.677 | 25.63% | 0.346 | 145.95% | 0.463 | 83.80% | 0.851 |
| JFreeChart | 0.833 | -13.21% | 0.774 | -6.64% | 0.340 | 112.65% | 0.481 | 50.31% | 0.723 |
| JMeter | 0.778 | 18.77% | 0.811 | 13.92% | 0.482 | 91.70% | 0.692 | 33.53% | 0.924 |
| JRuby | 0.750 | 21.47% | 0.818 | 11.32% | 0.560 | 62.68% | 0.718 | 26.88% | 0.911 |
| SQuirrel | 0.471 | 96.39% | 0.737 | 25.47% | 0.378 | 144.71% | 0.591 | 56.51% | 0.925 |
| Average | 0.729 | 21.84% | 0.747 | 18.81% | 0.420 | 111.23% | 0.557 | 59.50% | 0.888 |

Table 12. The performance comparison under the OTO scenario in terms of recall

| Projects | Pattern | Impr.% | NLP | Impr.% | TM | Impr.% | CNN | Impr.% | MAT |
|---|---|---|---|---|---|---|---|---|---|
| Ant | 0.098 | 370.41% | 0.256 | 80.08% | 0.456 | 1.10% | 0.359 | 28.41% | 0.461 |
| ArgoUML | 0.027 | 3355.56% | 0.745 | 25.23% | 0.768 | 21.48% | 0.544 | 71.51% | 0.933 |
| Columba | 0.070 | 1082.86% | 0.595 | 39.16% | 0.774 | 6.98% | 0.455 | 81.98% | 0.828 |
| EMF | 0.054 | 550.00% | 0.302 | 16.23% | 0.527 | -33.40% | 0.460 | -23.70% | 0.351 |
| Hiberbate | 0.072 | 905.56% | 0.608 | 19.08% | 0.704 | 2.84% | 0.587 | 23.34% | 0.724 |
| JEdit | 0.185 | 10.81% | 0.169 | 21.30% | 0.417 | -50.84% | 0.345 | -40.58% | 0.205 |
| JFreeChart | 0.050 | 1346.00% | 0.446 | 62.11% | 0.650 | 11.23% | 0.413 | 75.06% | 0.723 |
| JMeter | 0.050 | 1460.00% | 0.587 | 32.88% | 0.725 | 7.59% | 0.540 | 44.44% | 0.780 |
| JRuby | 0.055 | 1505.45% | 0.494 | 78.74% | 0.678 | 30.24% | 0.573 | 54.10% | 0.883 |
| SQuirrel | 0.040 | 1430.00% | 0.510 | 20.00% | 0.607 | 0.82% | 0.612 | 0.00% | 0.612 |
| Average | 0.070 | 827.25% | 0.471 | 37.95% | 0.631 | 3.08% | 0.489 | 32.98% | 0.650 |

Table 13. The performance comparison under the OTO scenario in terms of F$_1$

| Projects | Pattern | Impr.% | NLP | Impr.% | TM | Impr.% | CNN | Impr.% | MAT |
|---|---|---|---|---|---|---|---|---|---|
| Ant | 0.167 | 261.08% | 0.338 | 78.40% | 0.258 | 133.72% | 0.437 | 37.99% | 0.603 |
| ArgoUML | 0.052 | 1580.77% | 0.771 | 13.36% | 0.721 | 21.22% | 0.631 | 38.51% | 0.874 |
| Columba | 0.130 | 565.38% | 0.671 | 28.91% | 0.531 | 62.90% | 0.537 | 61.08% | 0.865 |
| EMF | 0.098 | 430.61% | 0.379 | 37.20% | 0.272 | 91.18% | 0.534 | -2.62% | 0.520 |
| Hiberbate | 0.133 | 516.54% | 0.719 | 14.05% | 0.656 | 25.00% | 0.622 | 31.83% | 0.820 |
| JEdit | 0.304 | 8.88% | 0.264 | 25.38% | 0.373 | -11.26% | 0.414 | -20.05% | 0.331 |
| JFreeChart | 0.093 | 677.42% | 0.546 | 32.42% | 0.436 | 65.83% | 0.487 | 48.46% | 0.723 |
| JMeter | 0.093 | 809.68% | 0.673 | 25.71% | 0.574 | 47.39% | 0.639 | 32.39% | 0.846 |
| JRuby | 0.102 | 779.41% | 0.607 | 47.78% | 0.606 | 48.02% | 0.667 | 34.48% | 0.897 |
| SQuirrel | 0.073 | 909.59% | 0.594 | 24.07% | 0.456 | 61.62% | 0.664 | 10.99% | 0.737 |
| Average | 0.125 | 479.92% | 0.556 | 29.74% | 0.488 | 47.78% | 0.563 | 28.13% | 0.722 |

Table 14. The results from the Wilcoxon-singed rank test and the Cliff's delta when comparing MAT with state-of-the-art approaches under the OTO scenario (bold fonts denote that the p-value is less than 0.05)

|  | Indicators | Precision | Recall | F1 |
|---|---|---|---|---|
| p-value | Pattern | **0.037** | **0.002** | **0.002** |
|  | NLP | **0.007** | **0.001** | **0.001** |
|  | TM | **0.002** | 0.359 | **0.004** |
|  | CNN | **0.002** | 0.058 | **0.014** |
| Cliff's delta (Effect size) | Pattern | 0.680 (large) | 1.000 (large) | 1.000 (large) |
|  | NLP | 0.785 (large) | 0.521 (large) | 0.570 (large) |
|  | TM | 1.000 (large) | 0.200 (small) | 0.680 (large) |
|  | CNN | 1.000 (large) | 0.510 (large) | 0.560 (large) |

Table 14 reports the statistical test results under the OTO prediction scenario. In the view of statistical test, we make the following observations. In terms of precision and F1, MAT is significantly better (p-value < 0.05) than all the supervised approaches (all the effect sizes are large). In terms of recall, MAT is significantly better than NLP (the effect size is large) and is similar to TM and CNN.

Combining the above results, we can see that, under the OTO scenario, the effectiveness of the state-of-the-art supervised approaches decline in various degree. Overall, MAT outperforms all the state-of-the-art approaches in identifying SATD comments under the scenario that only limited training data are available.

***Conclusion.*** *In summary, MAT exhibits a very competitive or even superior overall performance to the state-of-the-art approaches when identifying self-admitted technical debt.*

### 7.3 RQ3: Difference in correct classification results

*Result.* Fig. 7 reports the number of true positive instances (TPs) classified by NLP, TM, and MAT, while Fig. 8 reports the number of true negative instances (TNs) classified by NLP, TM, and MAT. In each figure, the blue, red, and green circles respectively denote the number of SATD comments that correctly classified by NLP, TM, and MAT. In addition, Table 15 reports the percentage of TPs and TNs classified by each approach.

**Table 15.** The percentage of true positive instances and true negative instances classified by each approach

| Projects | #TPs identified by three approaches | % of overlapped TPs | % of TPs identified by individual approach | | | #TNs identified by three approaches | % of overlapped TNs | % of TNs identified by individual approach | | |
|---|---|---|---|---|---|---|---|---|---|---|
| | | | NLP | TM | MAT | | | NLP | TM | MAT |
| Ant | 69 | 43.48% | 13.04% | 13.04% | 8.70% | 2947 | 97.32% | 0.03% | 0.00% | 0.31% |
| ArgoUML | 928 | 82.54% | 1.51% | 0.32% | 2.26% | 4323 | 96.53% | 0.44% | 0.46% | 0.16% |
| Columba | 114 | 70.18% | 1.75% | 0.00% | 1.75% | 3956 | 98.99% | 0.10% | 0.03% | 0.05% |
| EMF | 43 | 41.86% | 13.95% | 16.28% | 2.33% | 2511 | 97.73% | 0.00% | 0.00% | 0.04% |
| Hibernate | 305 | 70.49% | 3.28% | 3.61% | 0.66% | 2107 | 96.54% | 0.33% | 0.00% | 0.28% |
| JEdit | 97 | 24.74% | 18.56% | 16.49% | 1.03% | 4447 | 98.81% | 0.02% | 0.00% | 0.09% |
| JFreeChart | 81 | 67.90% | 4.94% | 1.23% | 0.00% | 2372 | 98.65% | 0.30% | 0.00% | 0.13% |
| JMeter | 233 | 78.54% | 2.58% | 1.72% | 1.29% | 3862 | 97.62% | 0.23% | 0.03% | 0.10% |
| JRuby | 359 | 67.41% | 2.79% | 0.56% | 5.01% | 3245 | 97.60% | 0.12% | 0.06% | 0.12% |
| SQuirrel | 149 | 69.13% | 9.40% | 2.01% | 2.01% | 4266 | 98.08% | 0.05% | 0.05% | 0.26% |
| Average | 237.8 | 61.63% | 7.18% | 5.53% | 2.50% | 3403.6 | 97.79% | 0.16% | 0.06% | 0.15% |
| Total | 2378 | 72.16% | 3.91% | 2.35% | 2.40% | 34036 | 97.84% | 0.16% | 0.08% | 0.15% |

According to Fig. 7, it can be seen that, the SATD comments correctly classified by NLP, TM, and MAT are largely overlapped. In total, there are 1716 (in 10 projects) common SATD comments correctly identified by NLP, TM, and MAT, which accounts for 83.46% of the SATD comments correctly identified by NLP (2056), 83.30% of the SATD comments correctly identified by TM (2060), and 79.81% of the SATD comments correctly identified by MAT (2150). This indicates that MAT can identify most SATD comments that NLP and TM can identify. Note that, the total number of SATD comments correctly identified by MAT is more than 94 of NLP and 90 of TM. It is interesting that such relationship cannot be reflected by the average recall (0.637 for NLP, 0.655 for TM, and 0.650 for MAT). In other words, the average recall can hide the fact that MAT performs the best in terms of the number of the correctly identified SATD comments.

Considering the number of SATD comments correctly identified by only one approach, NLP achieves 93 comments, TM achieves 56 comments, and MAT achieves 57 comments in all projects. In terms of the result of SATD comments in individual project, NLP and TM can identify more SATD in some projects (e.g., JEdit and EMF) while perform poor than MAT on other projects. For example, NLP and TM perform better in the project JEdit that they can identify 32 and 31 more SATD comments than MAT. The reason is that there are little task tags in the project so that MAT cannot identify many SATD comments by matching task tags. Meanwhile, MAT performs better than NLP and TM in the project JRuby since many SATD comments in this project have been marked by developers.

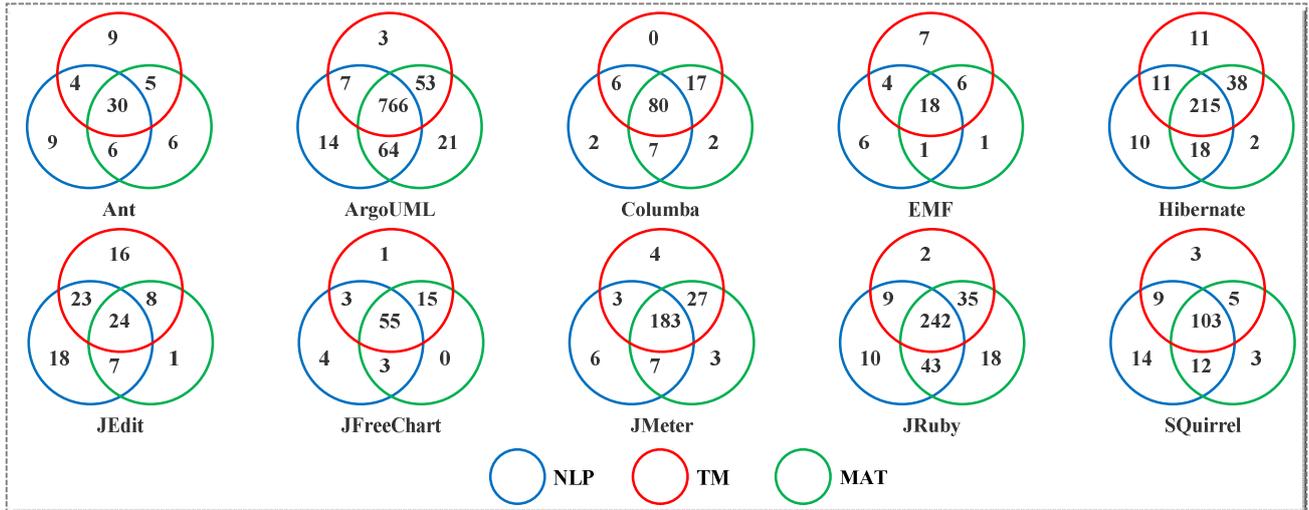

**Figure 7.** Set diagrams for the number of true positive instances classified by MAT and two supervised approaches (NLP and TM) in 10 projects

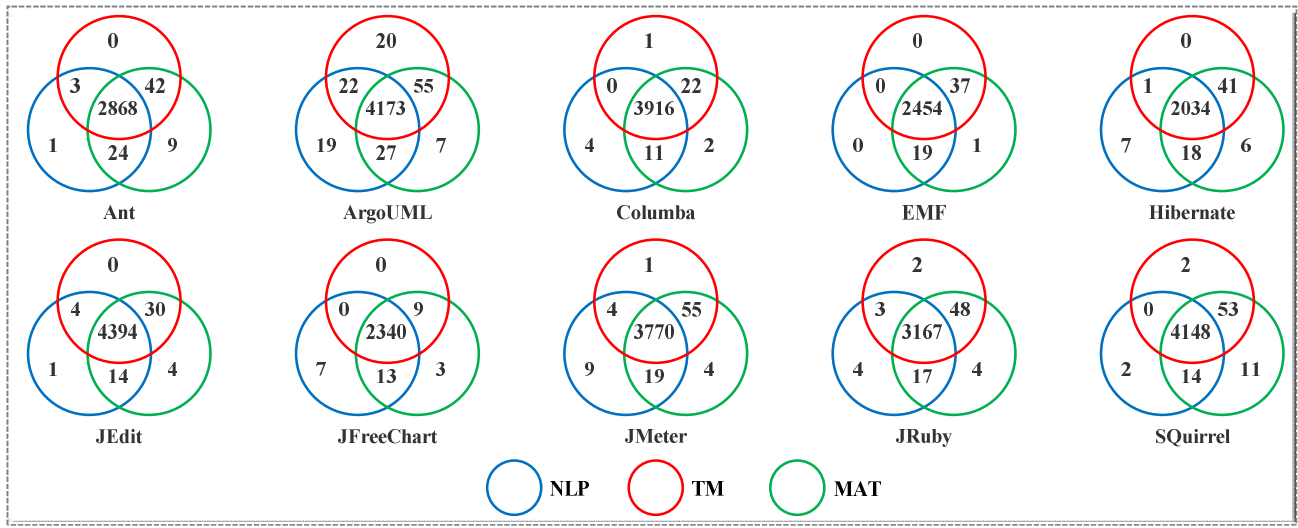

**Figure 8.** Set diagrams for the number of true negative instances classified by MAT and two supervised approaches (NLP and TM) in 10 projects

According to Fig. 8, it can be seen that, most of the non-SATD comments classified by NLP, TM, and MAT are also overlapped. In total, there are 33300 common non-SATD comments correctly classified by NLP, TM, and MAT, which accounts for 99.20% of the non-SATD comments identified by NLP (33567), 98.65% of the non-SATD comments identified by TM (33755), and 98.18% of the non-SATD comments identified by MAT (33919). This indicates that most non-SATD comments can be identified by all three approaches. In particular, we can see that MAT performs the best in terms of the number of the identified correct non-SATD comments. Compared with the overlapped non-SATD comments, the unique non-SATD comments identified by each approach are very little and hence there are no large difference among the three approaches.

*Conclusion. In summary, the SATD (non-SATD) comments correctly identified by MAT are largely the same as those correctly identified by the state-of-the-art supervised approaches. In addition, MAT can correctly identify more SATD (non-SATD) comments.*

# 8 Discussions

In this section, we first discuss important factors that may affect the effectiveness of MAT. Then, we analyze the limitations as well as the implications of MAT.

## 8.1 Can we extend MAT to achieve more accurate SATD identification by matching project-specific tags?

In the default MAT model, we match four popular task tags to identify SATD comments. Although it can achieve a promising performance in terms of average F1, we find that some projects achieve a low recall rate (e.g., 0.205 for JEdit and 0.351 for EMF) due to only few comments of the corresponding projects contain the four task tags. In this context, the following interesting problem is naturally raised: is there a simple approach to extend MAT such that the recall in such projects can be improved?

Table 16. The specified tags used in the subject projects

| Specified Task Tags | Projects | The meanings of tags | Examples |
|---|---|---|---|
| WORKAROUND | Columba Hibernate JEdit SQuirrel | Alternatives, the problem itself has not been solved | //**WORKAROUND**: we simply append URLs to the existing global class loader and use the same as parent - [from Columba] |
| TBD | EMF Hibernate | The abbreviation of "To Be Determined" | // **TBD** filter out volatile and other inappropriate links? - [from EMF] |
| REVISIT | EMF | There is something that need to be revisit | // **REVISIT:** Remove this code. // Store port value as string instead of integer. - [from EMF] |
| Note | JEdit | There is something that need to be noted | //"/* **Note:** This class is messy. The method and field resolution need to be rewritten. - [from JEdit] |
| NOTUSED | JMeter | The statements that not used | // **NOTUSED** private String chosenFile; - [from JMeter] |
| REMIND | JMeter | A weak code that should pay attention to later | // **REMIND:** convert arg list Vectors here? - [from JMeter] |

To tackle the above problem, we manually read the comments in these projects that have a low recall to understand their characteristics. Consequently, we make the following observations. First, the developers of these projects did not always use the default task tags (e.g., "TODO") in their comments. For instance, there are only 40 comments that contain default tags in all SATD comments (195) in project JEdit. Second, in addition to default task tags, there are project-specific task tags (defined by developers, usually highlighted in capitalized words) in these projects. For example, the word "NOTUSED" often appears in the comments in the project JMeter. Table 16 summarizes the meanings and examples of project-specific tags used in the subject projects under study. Intuitively, for a given project, if we adapt MAT to incorporate such project-specific task tags, the performance in SATD identification would be improved. For the simplicity of presentation, we name MAT incorporating project-specific tags as MAT-ext.

Table 17. The performance comparison between MAT and MAT-ext

| Projects that contain specified task tags | Precision | | | Recall | | | F1 | | |
|---|---|---|---|---|---|---|---|---|---|
| | MAT | MAT-ext | Impr% | MAT | MAT-ext | Impr% | MAT | MAT-ext | Impr% |
| Columba | 0.906 | 0.910 | 0.44% | 0.828 | 0.867 | **4.71%** | 0.865 | 0.888 | **2.66%** |
| EMF | 1.000 | 0.898 | -10.20% | 0.351 | 0.595 | **69.52%** | 0.520 | 0.715 | **37.50%** |
| Hibernate | 0.945 | 0.930 | -1.59% | 0.724 | 0.743 | **2.62%** | 0.820 | 0.826 | **0.73%** |
| JEdit | 0.851 | 0.683 | -19.74% | 0.205 | 0.441 | **115.12%** | 0.331 | 0.536 | **61.93%** |
| JMeter | 0.924 | 0.907 | -1.84% | 0.780 | 0.798 | **2.31%** | 0.846 | 0.849 | **0.35%** |
| SQuirrel | 0.925 | 0.923 | -0.22% | 0.612 | 0.652 | **6.54%** | 0.737 | 0.764 | **3.66%** |
| Average | 0.925 | 0.908 | -1.84% | 0.583 | 0.659 | **13.04%** | 0.687 | 0.748 | **8.88%** |

Table 17 reports the performance comparison between MAT and MAT-ext. As can be seen, on average, MAT-ext achieves an improvement of 13.04% and 8.88% compared with MAT in terms of recall and F1 in the six projects. This indicates that, the incorporation of project-specific task tags in MAT has a positive effect in identifying SATD comments. In particular, in project JEdit, the improvement of recall is 115.12%, since a large number of the tag "Note" exists in the comments of project JEdit. Fig. 9 reports the distribution comparison of F1 scores among CNN, MAT, and MAT-ext. We can see that, after incorporating project-specific task tags, MAT-ext achieves a more competitive performance compared with CNN. If we take into account the cost of modeling building and application, it is clearly that MAT-ext is preferred in practice.

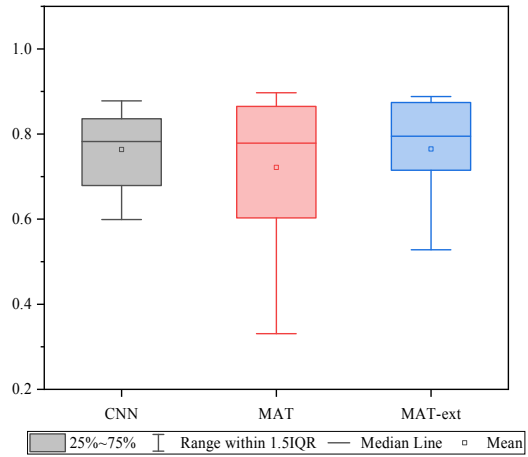

**Figure 9.** The distribution comparison of $F_1$ scores among CNN, MAT, and MAT-ext under the MTO scenario in 10 projects.

Indeed, for researchers, they do not know what tags are in a specific project in advance and only MAT can be applied in the experiments of research. However, for the developers that are responsible for the project, it is easy to acquire the project-specific task tags because developers in the project usually specify the used tags clearly at the stage of design in order to better maintain the project later. Therefore, it is meaningful to apply MAT-ext to improve the practicability of MAT in real projects.

**8.2 Can we achieve a more effective approach by combining the advantages of supervised approaches and MAT in SATD comments identification?**

According to the result of RQ1, MAT can achieve a high precision for each project. This means that the majority of comments that contain task tags are indeed SATD comments, which can be identified by MAT effectively and accurately. Meanwhile, according to the result of RQ2, there are 172/135 SATD comments that correctly classified by NLP/TM cannot be identified by MAT, since some SATD comments are not marked by task tags. In other words, the advantages of NLP/TM and MAT are different. This motivate us to combine MAT and NLP (or TM) to achieve a more excellent classification result.

To this end, for each target project, we first apply MAT to identify the comments that contain task tags and predict them as SATD comments. Then, we apply NLP (or TM) to classify the remaining comments that do not contain any task tags into two categories: SATD and non-SATD. Finally, we combine the SATD comments predicted by two approaches (MAT and NLP or MAT and TM). For easy of presentation, we name the two combined approaches NLP+MAT and TM+MAT respectively.

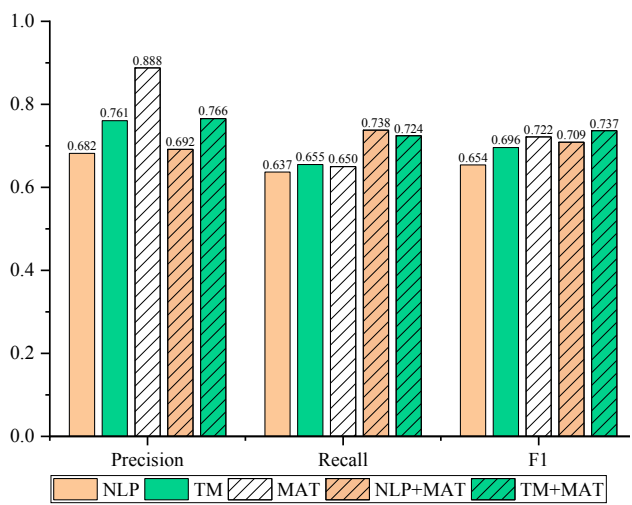

**Figure 10.** The average performance comparison among NLP, TM, MAT, NLP+MAT, and TM+MAT

Fig. 10 reports the average performance comparison among NLP, TM, MAT, NLP+MAT, and TM+MAT. As can be seen, the performances of two combined approaches (NLP+MAT and TM+MAT) are higher than the original supervised approaches (NLP and TM) respectively. More specifically, the recall and F1 of NLP+MAT have an improvement of 15.86% and 8.41% respectively compared with NLP. TM+MAT achieves a performance of 0.724 and 0.737 in terms of recall and F1. The recall has been improved by 10.53% and the F1 has been improved by 5.89% compared with TM, respectively. The results indicate that MAT has a positive effect for recall and F1 when it is combined with the supervised approaches (NLP or TM). Meanwhile, the precision values of two combined approaches are not affected negatively.

Compared with MAT, the two combined approaches both improve the recall of MAT largely in the average percentage of 13.54% (NLP+MAT) and 11.38% (TM+MAT) respectively. In particular, the improvements of recall are large for projects Ant, EMF, and JEdit. For example, TM+MAT achieves an improvement of 27.55%, 42.45%, and 97.56% compared with MAT in the above three projects respectively. These improvements indicate that the combined approaches can perform better for recalling more SATD comments that do not contain task tags. Nevertheless, the precision of two combined approaches are lower than that of MAT since the NLP (or TM) will misclassify some non-SATD comments. This means that some false positive comments are introduced.

On the whole, the combined approaches (NLP+MAT or TM+MAT) have a better performance than single supervised approaches (NLP or TM). By combining NLP (or TM) and MAT, more SATD comments are identified. Therefore, it is recommended to use when there are little task tags in the comments of one project.

**8.3 What are the limitations when applying MAT?**

Although the excellent performance of MAT has been shown in Section 7, there are some inherent limitations when applying MAT to identify SATD comments due to the simple design of MAT, i.e., MAT can output some mis-classified comments. According to our thoroughly analysis, we conclude the mis-classification of MAT into the following four kinds of situations. In addition, we report the summary of mis-classification of MAT in Table 18 to illustrate the limitations of MAT.

Table 18. The summary of mis-classification of MAT

| Type | Reasons of mis-classification | Examples |
|---|---|---|
| False Positive | The FP caused by the auto-generated tag comments. | // **TODO**: - [from ArgoUML]<br>//**TODO**!!! - [from Hibernate]<br>// **FIXME** - [from JFreeChart]<br>/* **XXX***/ - [from JRuby] |
| | The FP caused by the components of sentence. | /* Owner related **todo** items: */ - [from ArgoUML]<br>// Copy the **todo** items after the model - [from ArgoUML]<br>// no item exists in table // -> nothing **todo** - [from Columba]<br>// **Hack** to ensure charset is set correctly at start-up - [from Columba] |
| False Negative | The FN caused by the untagged comments. | // Check it out; also ugly. - [from Ant]<br>// Our superclass no longer has this method - [from ArgoUML]<br>// this part sucks - [from JEdit]<br>// Remember to change this when the class changes ... - [from JMeter]<br>// Not implemented - [from SQuirrel] |
| | The FN caused by the forms of tags. | // **TO DO**: these annotations only work with XYPlot - [from JFreeChart]<br>//**TO DO** : delete the file if it is not a valid file. - [from Ant] |

**1) The false positive instances caused by auto-generated tag comments.** For the IDEs (e.g., Eclipse) that integrate the function of tagging comments, developers may introduce an auto-generated comment, which consists of only a task tag, due to their negligence. For example, there are some comments like "/* XXX*/" in JRuby (in Table 13) generated by IDE. However, it is unclear whether a developer tends to use such a comment to indicate SATD. Therefore, it is uncertain that whether these comments indicate SATD or not. In the dataset provided by Maldonado et al. [17], this kind of comments is manually classified as non-SATD. Note that, MAT will classify these comments into SATD comments since task tags occurs in them. This will introduce false positive instances.

**2) The false positive instances caused by the components of sentence.** According to our observation, there is another kind of false positive instances: some non-SATD comments contain the same word (e.g., "hack") as a task tag but it is not tended to be used as a tag. For the example of "/* Owner related todo items: */" from the project ArgoUML, the word "todo" is not a task tag but a component word of the comment sentence. Note that, these comments are not marked by task tags. The reason of the mis-classification of MAT is that task tags (e.g.,

"TODO") will be transformed into its original form (e.g., "todo") in the preprocessing stage before matching. This will be confused with the components of sentence and hence cause false positive instances.

**3) The false negative instances caused by the untagged comments.** This is the main kind of mis-classification of MAT. The rationale of MAT is based on the assumption that developers will use tags to mark the technical debts admitted by themselves. However, this is not the habit of some developers and hence MAT will perform poor in the projects for which these developers are responsible. For these comments, we can use the state-of-the-art supervised approaches (e.g., TM [10] and CNN [49]) to identify the SATD comments by extracting useful semantic information from comments.

**4) The false negative instances caused by the formats of tags.** This kind of mis-classifications are uncommon but also shows the weakness of MAT. In the application of MAT, we have considered to use fuzzy matching strategy to avoid some mis-classification. However, fuzzy matching we used cannot handle all forms of tags. For example, for the SATD comment "//TO DO : delete the file if it is not a valid file" from project Ant, MAT will classify it as a non-SATD comment because there is no task tag after preprocessing ("TO DO " will be transformed as two words "to" and "do" but not "todo"). This problem could be tackled by enhancing the fuzzy strategy in MAT.

It should be pointed out that, although there are the above limitations, they do not have a large influence on the overall effectiveness of MAT. The reason is that the percentage of mis-classifications is low when considering the large number of comments.

## 8.4 The implications of MAT

Our work has important implications for both practitioners and researchers.

- For practitioners, we provide an easy-to-use approach MAT to identify SATD. Being a simple unsupervised approach, MAT is training-free, does not involve any parameter-tuning, and has a very low computation cost. Furthermore, MAT is very competitive or even superior to the-state-of-the-art approaches in SATD identification. In particular, if needed, MAT can be easily enhanced by incorporating project-specific task tags. These advantages make MAT efficient and effective in SATD identification.

- For researchers, we contribute a simple yet strong baseline in SATD identification. In the future, if a new identification approach is proposed, it should be compared against MAT to demonstrate its practical value. This will help filter out actually useless approaches and find really effective approaches. What is more important, our work demonstrates an important software engineering scenario that a simple solution could work well compared with complex solutions. Indeed, in software engineering, it is not uncommon to observe similar phenomena. For example, in cross-project defect prediction, we found that, ManualDown, a very simple module size model, performed similarly to or even better than the state-of-the-art (complex) supervised models [64]; in the Stack Overflow text mining task, Majumder et al. reported that a simple tuned local model performed similarly to the state-of-the-art CNN model but was 500+ times faster [77]; in the automatic generation of commit message, Liu et al. showed that a simple nearest neighbor generator outperformed the complex neural machine translation algorithm but was 2600 times faster [75]. The above facts caution that when faced with a software engineering problem, researchers should first seek simple rather than intricate and complex solutions. This is in accordance with the idea of "less, but better" advocated by Menzies [76], which aims to use simple approaches to solve complex problems. At least, "before researchers release research results, they compare their supposedly sophisticated methods against simpler alternatives" [77]. This will help avoid wasting research effort and find practical solutions.

# 9 Related Works

We first introduce the related studies on source code comments in software quality management and technical debt (especially self-admitted technical debt, SATD). Then, we present the application of unsupervised models in software engineering.

## 9.1 Source code comments

Source code comments play an important role in the field of software quality management [27, 36, 38-40]. The purpose of writing comments is to make code easy to understand and manage. Code comments are usually written in natural language text to assist project development (e.g., functional description and task annotation). Therefore, we can extract valuable information from comments using a natural language process technology to solve software quality problems.

Many studies investigated the characteristics of source code comments. Fluri et al. [27] investigated to the level of developers added comments or adapted them when they evolved the code in three open source systems (e.g., ArgoUML). Consequently, they found that, as the associated code change, comment changes often were done in the same version. Malik et al. [28] used a random forest model to predict the likelihood of a comment being updated and reported. Results in four large open source projects showed that this model had a high accuracy (i.e., 80%). Tan et al. [34] and Khamis et al. [35] studied a special kind of comments (i.e., Javadoc comments). Tan et al. found 29 inconsistencies between Javadoc comments and method bodies. Note that, five of them had been confirmed and fixed by developers. Khamis et al. presented JavadocMiner to automatically assess the quality of inline documentations, including quality of language and consistency between source code and its comments. Steidl et al. [37] and Sun et al. [41] conducted source code comment quality analysis to assess the quality of code comments. More specifically, Steidl et al. provided a semi-automatic approach to conduct quantitative and qualitative evaluation of comment quality. Sun et al. extended their work and made a more accurate and comprehensive comment assessment and improvement recommendation.

## 9.2 Technical debt

Technical debt will reduce the quality of software and increase the cost of software maintenance. Therefore, a large number of studies [12-13, 23-26, and 29-32] were devoted to explore the characteristics of technical debt and the identification of technical debt. Zazworka et al. [19] conducted a case study to investigate the impact of design debt on software quality. They found that technical debt could negatively affect the software quality so they suggested that developers need to identify and manage technical debt closely in the software evolution process. Tom et al. explored technical debt in order to understand the nature and the implications of technical debt. They proposed a framework that provided a useful approach to understanding the overall phenomenon of technical debt for practical purposes. Furthermore, Li et al. [20] performed a systematic mapping study on technical debts and their management. They concluded that there was a need for more empirical studies with high-quality evidence on the whole technical debt management (TDM) process and on the application of specific TDM approaches in industrial settings. Digkas et al. [45] studied the characteristics of technical debt in the apache ecosystem. One of their findings was that, for the most frequent and time-consuming types of technical debt, some of which were related to inappropriate exception handling and code replication. In addition, a small subset of all issue types was responsible for the largest percentage of TD repayment. As a result, by targeting particular violations, the development team could achieve higher benefits.

Recently, a kind of technical debt called self-admitted technical debt (SATD) proposed by Potdar et al. SATD denotes that technical debt is introduced intentionally for achieving short-term goals. In order to facilitate the repayment of SATD, developers usually write comments in the nearby of the corresponding source code. The existing studies [23, 24] show that SATD is important since they are common in software project. What is more, it has a negative impact on software quality. Currently, a number of approaches have been proposed to identify SATD. Potdar et al. [14] proposed to use comment to identify self-admitted technical debt. Subsequently, many researchers conducted research on the field to study SATD according to code comment. Wehaibi et al. [15] conducted an experiment to examine the impact of SATD on software quality, especially on software defects. Bavota et al. [16] conducted a differentiated replication of Potdar et al.'s work [14] to investigate the diffusion and evolution of SATD and its relationship with software quality.

## 9.3 Task tags

Task tag is a special kind of code comment entity, which can be used to manage and coordinate programming tasks. Ying et al. [59] conducted an informal empirical study to research the use Eclipse task comments in Java source code. According to their study, task comments (i.e., the comment marked by task tags) contained many useful information and they were intended for many purposes. For example, developers used the task comments as a medium to communicate to each other. In addition, some comments that contained task tags denoted outstanding tasks that needed to be done currently or in the future. Storey et al. [22] also conducted an empirical study to explore how task annotations played a role in the work practices for software developers. According to their research, task annotation tags were important for individual, team, and community use in software maintenance. Subsequently, they examined how software developers used task tags to help them remember pertinent information and mark locations of interest for future investigation [58]. Based on this, they designed a new approach (i.e., TagSEA) to navigate these tagged comments. Their tool provides support for creating, editing, navigating, and managing these tags. Their findings also showed that the semantic information can improve the value of annotations tags.

## 9.4 Unsupervised models in software engineering

Similar to the idea in this paper, there have been large number of studies on unsupervised models in different fields of software engineering [53-57]. One advantage of unsupervised models is that their construction does not require the label of instances. Therefore, these unsupervised models can be very effective when the labels of instances are difficult to acquire. What is more, the computational cost of unsupervised models usually is low compared with supervised models. This also motivates researchers to study unsupervised models in the field of software engineering.

Yang et al. [53] applied unsupervised models in the issue of effort-aware just-in-time (JIT) defect prediction. They investigated the predictive power of the simple unsupervised models (e.g., LT) in effort-aware JIT defect prediction and compared them with the state-of-the-art supervised models under different prediction settings (including cross-validation, time-wise-cross-validation, and across-project prediction). According to their result, the unsupervised models performed better than the supervised model in effort-aware JIT defect prediction. Huang et al. also conducted a holistic look at effort-aware JIT defect prediction [56] based on the Yang et al.'s work. They presented new findings such as the false alarms of LT are higher [55, 56]. In addition, Zhou et al. [64] investigated how far researchers had really progressed in the journey. They compared the performance of simple unsupervised module size models with existing cross-project defect prediction (CPDP) models. As a result, they found that ManualDown had a comparable or even superior prediction performance compared with the existing state-of-the-art CPDP models.

Fu et al. [54] conducted a case study on deep learning and proposed to use an unsupervised model to solve the problem of which questions in the Stack Overflow programmer discussion forum can be linked together. According to their results, their model can achieve similar (or even better) results compared with the supervised deep learning model. Meanwhile, their model can reduce the cost of total time of deep learning model by 84 times. Following up Fu et al.'s work, Xu et al. [57] replicated the previous studies and further investigated the validity of Fu et al. claimed by evaluating both supervised and unsupervised based approaches on a larger dataset. Meanwhile, they also compared the effectiveness of the existing approaches against a lightweight SVM-based method (SimBow) that was previously used for general community question-answering. Their findings showed that the performance of unsupervised based approaches (Fu et al.'s model and SimBow) were slightly better than the supervised model. In addition, SimBow outperformed both supervised approach and Fu et al.'s approach when considering the runtime.

## 10 Threats to Validity

We consider the most important threats to construct, internal, and external validity of our study.

### 10.1 Construct validity

Construct validity is the degree to which the dependent and independent variables accurately measure the concept they purport to measure. In this study, the used dependent variable is a binary variable that indicates whether a code comment is SATD comment. We used the dataset shared online by Huang et al, which was collected by Maldonado et al. [17] According to Maldonado et al.'s study, they manually read comments and then labeled the category to obtain the dependent variable. Although they had double inspected the dataset, there may be also ambiguous comments that cannot identify the classification exactly. For example, the comments that only consists of a task tag (e.g., "// TODO:") cannot be determined whether these comments indicate SATD. Therefore, this is a threat to the construct validity of the dependent variable that needs to be reduced in the future work.

**10.2 Internal validity**

Internal validity is the degree to which conclusions can be drawn about the causal effect of independent variables on the dependent variables. The most important threat is from the implementation of fuzzy matching strategy in MAT. As mentioned, MAT uses a fuzzy matching strategy to find the four tags in code comments. In our implementation, we consider two words will be linked together due to developers' negligence (delete spaces). Therefore, we think a comment will indicate SATD if the head or tail of one of its tokens match the tags (e.g., "pleasefixme" will match "FIXME"). However, such a simple implementation may miss some tokens indicating SATD (e.g., "TO DO", "to-do" and "tod" cannot match "todo"). This threat needs to be reduced in the future work.

**10.3 External validity**

External validity is the degree to which the results of the research can be generalized to the population under study and other research settings. The most important threat is that our findings may not be generalized to non-Java projects or commercial projects. In our experiment, we use the dataset shared online by Huang et al., in which all the projects are open-source Java projects. In particular, MAT uses four tags (i.e. "TODO", "FIXME", "HACK", and "XXX") to identify SATD. For non-Java or commercial projects, if the tags for marking SATD are different from these four tags, MAT will not work. One possible solution is to adapt MAT by replacing the four tags with the popular tags or developer-defined tags in those projects. In this context, we believe that the adapted MAT should still work. Nonetheless, in order to mitigate this threat, there is a need to replicate our study across a wider variety of projects in the future work.

# 11 Conclusion and Future Work

In this paper, we propose a simple approach called MAT to identify SATD by matching task tags in source code comments. In nature, MAT is an unsupervised approach, which does not need the training data to train a prediction model. Based on 10 different open-source projects, our experimental results show that, in terms of classification effectiveness, MAT is very competitive or even superior to all the state-of-the-art approaches. Furthermore, MAT has a low computation cost and a low memory requirement. Therefore, MAT can be effectively and efficiently applied in practice. As such, we strongly recommend that future SATD identification studies should consider MAT as an easy-to-implement baseline. In particular, in the light of the fact that task tags are usually signals of SATD, there is no reason to neglect such a natural baseline. In summary, using MAT as a baseline will enable us to determine whether a new SATD identification approach is practically useful.

In the future, we will collect more comments from other projects to evaluate the effectiveness of MAT. What is more, we will investigate developers' habit of writing comments to mine more information about SATD and provide a more effective identification approach.

## Acknowledgment

We are very grateful to Q. Huang, E. Shihab, X. Xia, D. Lo, and S. Li for sharing their comment data sets and TM-based source code online. In addition, we are very grateful to Ren for providing the details on the data sets used in their CNN-based approach.
## Acknowledgment

We are very grateful to Q. Huang, E. Shihab, X. Xia, D. Lo, and S. Li for sharing their comment data sets and TM-based source code online. In addition, we are very grateful to Ren for providing the details on the data sets used in their CNN-based approach.